\documentclass[manuscript,nonacm]{acmart}
\AtBeginDocument{%
  }

\usepackage{todonotes}
\usepackage{amsmath}
\usepackage{pdfpages}
\usepackage{booktabs}
\usepackage{pifont}
\usepackage{tabularx}
\usepackage{ragged2e}
\usepackage{balance}
\usepackage{multirow}
\usepackage{enumitem}
\usepackage{graphicx}
\usepackage{subcaption}
\usepackage{geometry}
\usepackage{copyrightbox}
\usepackage{url}
\usepackage{stackengine}
\usepackage{longtable}
\newcommand{\Planner}{\textsc{pragmatic planner}}

\newcommand{\PlannerCap}{\textsc{Pragmatic Planner}}
\newcommand{\Baseline}{\textsc{baseline}}
\newcommand{\BaselineCap}{\textsc{Baseline}}
\newcommand{\PP}{\textsc{pp}}

\begin{document}

\title[Must AI Spell It Out?]{Must AI Spell It Out? Pragmatic Comprehension of Alerts in a Time-Critical Task}

\author{Anwesha Das}
\email{adas@lst.uni-saarland.de}
\orcid{0009-0006-8308-9961}
\affiliation{%
 \institution{Saarland Informatics Campus, Saarland University}
 \city{Saarbrücken}
 \country{Germany}
}

\author{Vera Demberg}
\email{vera@lst.uni-saarland.de}
\orcid{0000-0002-8834-0020}
\affiliation{%
 \institution{Saarland Informatics Campus, Saarland University}
 \city{Saarbrücken}
 \country{Germany}
 \postcode{66123}
}

\author{Anna Maria Feit}
\email{feit@cs.uni-saarland.de}
\orcid{0000-0003-4168-6099}
\affiliation{%
 \institution{Saarland Informatics Campus, Saarland University}
 \city{Saarbrücken}
 \country{Germany}
}

\renewcommand{\shortauthors}{Das et al.}

\begin{abstract}
  Assistive systems typically tell users everything, often spending time and attention on what a user already knows, leaving less for what happens next. People, on the other hand, leave out what they expect the listener can supply. Can people use their prior knowledge to understand an incomplete message when the speaker is a machine though? We studied this pragmatic comprehension in a monitoring task controlling users' prior knowledge and the available time to detect critical situations. In a preregistered study ($N=114$), we compared a \Planner{} that reasons over a model of the listener's knowledge to decide what to say, when, and how explicitly, with a \Baseline{} specifying problems fully in onset order. Participants resolved underspecified alerts almost as reliably as fully specified ones and sooner, detected more problems, and rated the pragmatic assistant helpful more often, with no increase in reported mental demand.
  Our findings show that people can recover what the system left implicit and that pragmatic communication can support effective assistance under the constraints of a changing world.
\end{abstract}

\begin{teaserfigure}
  \centering
  \includegraphics[width=\textwidth]{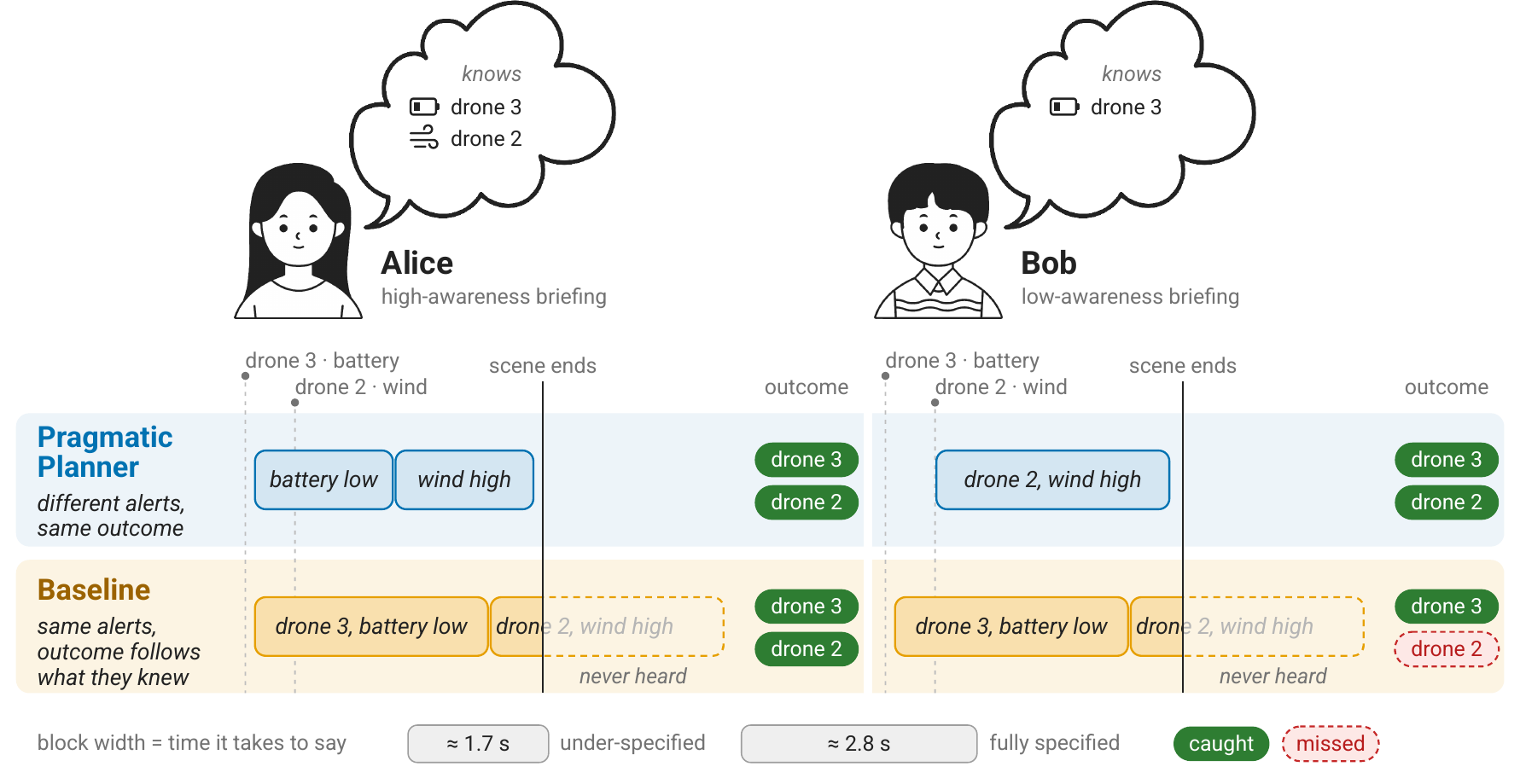}
  \caption{
  Typical alerting systems name every problem in full in the order they occur. People instead reason about what their listener knows and state only what they cannot supply. In a user study ($N=114$), we examine whether people can infer what an assistant leaves underspecified when it communicates in this pragmatic way. The figure shows one scene reaching two people who know different things: Alice was briefed on both problems to come, Bob only on the first. The \Baseline{} (bottom) alerts both identically and runs out of time one word into its second alert, leaving Bob unaware of drone 2. The pragmatic assistant (top) omits what each can supply and spends the time saved on what they cannot, so Bob catches both.
  }
  \label{fig:1}
\end{teaserfigure}

\maketitle

\section{Introduction}
Your assistive systems are built to tell you everything. Ask a navigation system about the way and it gives the same message no matter if you are a visitor or a local who has lived on that street for twenty years. That habit of completeness runs through even to assistive systems in high-risk or time-critical tasks. A surgeon needs decision-support mid-procedure~\cite{ancker2017effects}, a nurse tracks the alarms that cross an intensive-care ward~\cite{gulcsen2025effect}, an operator needs time-critical warnings about the systems they monitor~\cite{khan_exploratory_2019,wu2025releyeance}. In each case, systems give full information to their users as soon as a critical situation occurs~\cite{joseph2021alert}.
This can be costly in time-critical and high-risk scenarios.
Operators who spend limited time and attention on what they already know risk missing critical situations that come next~\cite{lieder2026rational,horvitz1999attention}.
An assistant's overspecification without regard to their human partner then becomes expensive, and the resulting harm is well documented~\cite{ancker2017effects,gulcsen2025effect,mongan2020artificial, joseph2021alert}.

People, by contrast, when communicating with one another, adapt messages to their partner constantly and, without effort, avoid overspecifying. Ask the way in a town you grew up in and your pragmatic friend will tell you \textit{``Left at the old cinema.''}, where they would tell a visitor \textit{``Straight on through the traffic lights, then the left just after the mall's petrol station.''} They describe the same route, but reason about the listener's knowledge to say only what they believe the listener cannot themselves supply~\cite{grice1975logic,hawkins2021}. This is a rational and economical use of our limited
attention~\cite{lieder2026rational}, and pragmatic speakers do it systematically~\cite{piantadosi2012communicative,kanwal2017language,tourtouri2021rational}. The Rational Speech Act (RSA) framework proposes a computational account of this behaviour. A speaker chooses an utterance by simulating how a listener with given prior beliefs will interpret it~\cite{frank2012,goodman2016,degen2023,franke2011,goodman2013}.
Research shows that language models that incorporate this reasoning describe, refer and instruct more accurately~\cite{monroe2017colors,andreas2016reasoning,fried2017unified}. In contrast, consider generative systems, which, for all their competence, still systematically over-specify~\cite{tang2024,kapur2025} and produce misunderstandings more often than human speakers do~\cite{tang2024}.

Recent work has brought this pragmatic reasoning to assistive alerting settings, through an RSA-based speaker model that plans its message sequences to adapt to its listeners~\cite{das2026planning}. It tracks what a user believes as the world changes and makes use of ambiguous, underspecified alerts wherever it judges the user can complete the reference, spending the time and resources saved on the next alert~\cite{das2026planning}. However, whether people effectively interpret and cooperate with such a pragmatic system has not been tested.

This is one instance of a broader argument that for effective cooperation with humans, systems ought to reason over a model of their human partner, and treat their user as an active partner, so that cooperation is designed in rather than hoped to emerge~\cite{gmytrasiewicz2001rational,amershi2019guidelines,collins_building_2024,miehling_language_2024,zhi2024pragmatic,saad2025gricean,oulasvirta2022computational}. However, this argument rests on an assumption that has drawn far less investigation than the machine side of it: that the user will, in return, reason over the system and complete what it leaves unspecified. Evidence in either direction is thin. While linguistics studies using pragmatic speakers in constrained settings showed promise~\cite{andreas2016reasoning,fried2017unified}, studies with people suggest they might not even attempt the inference with a computer partner~\cite{duran2011listeners} at all. Despite rating ChatGPT as highly capable, people interpreted the same message less pragmatically when told that it came from ChatGPT rather than a person~\cite{mayn2024capable}. Crucially, none of the evidence was gathered under conditions that assistive alerting demands. In no study so far, was a human listener placed in a dynamic environment with a time-critical task to test if they could produce gains through pragmatic inference.

In this paper we aim to empirically establish whether people comprehend messages from an assistant that communicates pragmatically  and whether such assistance helps in time-critical tasks.
For this we introduce DroneWorld, a dynamic monitoring task inspired by earlier work~\cite{das2026planning, wu2025releyeance}. In a preregistered study, participants ($N=114$) oversaw four drones on which problems turned critical in close succession, their task being to detect each as it arose. Two sources of information supported them: their prior knowledge and the assistant's messages. A briefing before each scene covered most (high awareness) or few (low awareness) of the problems to come, so that each participant was in turn Alice and Bob of Fig.~\ref{fig:1}. Between participants, the assistant was either the \Planner{} of~\citet{das2026planning}, which varies the specificity of its spoken alerts by reasoning about the user's prior knowledge, or a \Baseline{} announcing every problem in full, similar to typical alerting systems~\cite{ancker2017effects,joseph2021alert}. The world advanced whether or not the assistant spoke or the participant acted.

We found that participants were highly effective in using their prior knowledge to interpret the \Planner{}'s underspecified alerts. They resolved these almost as reliably as fully specified messages and detected relevant problems sooner after alert onset, leveraging the time gained from shorter messages. Across the task, participants detected more problems with the \Planner{}, and sooner, rating its assistance as helpful more often, with no evidence of greater reported mental demand or frustration. Interestingly, the largest performance advantage occurred in low-awareness scenes, where participants had been briefed about fewer forthcoming problems and where the planner brought more previously unbriefed problems to users' attention.

To summarise, we contribute: $(1)$ \emph{the first human evaluation of a pragmatic assistance system in a dynamic environment}, providing empirical evidence that people can perform pragmatic comprehension with an assistive system in a dynamic, time-critical environment, and an account of how this capability relates to the accuracy, timeliness and practical value of assistance systems. As part of our work, we publish $(2)$ \emph{DroneWorld}, a reusable environment for studying dynamic reference resolution in pragmatics and monitoring tasks, including a tested approach for manipulating users' awareness and a simulation of a time-critical task. $(3)$ \textit{More generally}, our work supports the call for interactive systems that reason over a model of their human partner, showing that the cooperation such systems assume is reciprocated, and that a system which trusts its user to reason about what it underspecifies or leaves unsaid is not only more efficient but also preferred.

\section{Background and Related Work}\label{sec:background}

We first introduce pragmatics in human--human communication and the RSA framework. We then discuss assistive systems that model their users, and evidence about how people interpret communication from pragmatic systems.

\subsection{Pragmatic Speakers and Their Listeners}\label{sec:RW_RSA}
Listeners use prior knowledge to understand more than an utterance (message) states literally~\cite{grice1975logic}. The Rational Speech Act (RSA) framework formalises this understanding as Bayesian inference~\cite{frank2012,goodman2016,degen2023} over a speaker's intended meaning. We call this use of context and prior knowledge to recover a speaker's intended meaning~\cite{sperberwilson1995,clark1996} \textit{pragmatic comprehension}, and use it to mean the human capability of prior-guided disambiguation throughout this paper.
 In reference games, a listener $L$ hearing an utterance $u$ infers its intended referent property $p$ by combining the likelihood that a speaker would choose $u$ to refer to $p$, $S(u \mid p)$, with prior beliefs about that referent, $P(p)$ as:
\[
L(p \mid u) \propto S(u \mid p)\,P(p).
\]
An utterance meaning several referents can therefore support different interpretations for listeners with different priors. Where that prior $P(p)$ strongly favours one referent, the listener can resolve it from the same message $u$ that would leave another listener uncertain. RSA also explains scalar implicature: why \textit{some} means \textit{not all}~\cite{franke2011,goodman2013} and reference resolution: how \textit{the red one} picks out a car in a crowded street~\cite{frank2012,monroe2017colors,hawkins2021}.

A pragmatic speaker $S$, in turn, uses this modelled listener $L(p\mid u)$ to choose what to say, preferring the $u$ with which the listener would most likely understand the intended $p$, weighing informativeness against the utterance's cost $C(u)$. A common RSA utility~\cite{goodman2016,scontras2021practical} expresses this trade-off as:
\[ S(u \mid p) \propto \log L(p \mid u) - C(u).\] Specificity follows: where a short, underspecified $u$ leaves them nearly as certain, its lower cost can make it preferable. When priors $P(p)$ do not leave several interpretations plausible, additional specification becomes more valuable. Models that describe, refer and instruct~\cite{monroe2017colors,andreas2016reasoning,fried2017unified} put into effect this capability of pragmatic communication, and report of its advantageousness. This adaptation depends on the speaker's estimate of the listener's knowledge.

RSA in its classical form settles a single utterance in a world that holds still. Only lately have these models incorporated dynamics at all, planning over sequences and deadlines rather than single utterances~\cite{das2026planning}; in previous evaluations, the interaction was untimed, turn-based or a single reference task~\cite{andreas2016reasoning,fried2017unified}. Some extensions carry the listener's belief forward in time, modelling word by word~\cite{cohngordon2019incremental}, or across conversational turns~\cite{anderson2021}. The \Planner{} used here combines listener modelling with planning over successive alerts in a changing environment, within dynamic RSA (d-RSA)~\cite{das2026planning}. It has the three properties our question requires: a model of the listener's beliefs that updates as the world evolves, alerts of variable specificity, and a plan over a sequence of them under a fixed speaking budget. It predicts how alerts direct the listener's attention to properties of the environment, how attending to those properties updates their beliefs, and how those beliefs inform subsequent interpretation. It then selects an alert sequence that maximises the expected accuracy of those beliefs about currently critical properties over a fixed horizon. Each alert occupies the speaking channel for its duration, so greater specificity can delay another message. The trade-off between informativeness and cost thus extends across time: saying less about a problem the listener can resolve leaves time to communicate another that needs more explicit support. Against such listeners, the models predict a gain, largest for the listener who already knows most and is therefore best placed to complete what is omitted~\cite{das2026planning}.

All of this rests on one assumption about the listener: that they run the matching inference, filling what the speaker leaves unsaid from their own priors, and keep doing so as the world moves on. Pragmatics has largely been free to assume it. Its evidence comes from exchanges between people, and the speaker models that plan around an evolving listener have so far been tested only against simulated ones~\cite{das2026planning,estienne2025collaborative}, listeners that complete the reference on cue and never wonder whether the speaker has simply overlooked something. Whether a real person does the same in a dynamic, time-critical setting, where their attention is already spent and an alert that says too little competes with the next event, is untested. We take up this question, and the further twist that arises once the speaker is a machine, in Sec.~\ref{sec:pragmaticsystems} below.

\subsection{Assistive Systems that Model Their Users}\label{sec:RW_HCI}

People's knowledge and attention have long informed user-modelling for adaptive systems in HCI~\cite{fischer2001user}. Mixed-initiative interfaces weigh an intervention's expected benefit against its cost to the user~\cite{horvitz1999principles}; attention-sensitive alerting integrates the costs of interruption and delay~\cite{horvitz1999attention}. Model-based interfaces adapt layout and presentation to users' abilities, tasks, or context~\cite{gajos2004supple,todi2021adapting,lindlbauer2019context}. Even in dialogue, referring expressions have been tailored to a user model updated over the course of interaction~\cite{janarthanam2014adaptive}. In each case, the model serves the system's decision about what to do \emph{for} the user. What the user will do with the result, and in particular what they can infer that the system need not state, is left outside the model. The human is treated as a recipient, not a partner.

The theory of computational rationality has provided the theoretical counterpart on the user's side: it explains a user's interactive behaviour as the optimal use of their limited perceptual, cognitive, and motor resources under the constraints of the task and the interface~\cite{howes2009rational,lewis2014computational,lieder2020resource,oulasvirta2022computational}. Models built on this premise can predict human behaviour in a variety of domains from typing to driving~\cite{jokinen2021touchscreen,howes2026,li2026simulatingwordsuggestionusage,sun2025adaptive}. While they can be used to generate behaviour, they mostly describe what a user does, not what a user and a system do together. Modelling assistance requires considering both sides at once. \citet{gori2026decision} make this relationship explicit by modelling assistance as a two-agent game in which the assistant benefits from tracking what the user believes, for instance, anticipating that a user missing a piece of information will act suboptimally. From the assistant’s perspective, this can be reduced to a single-agent problem in which the user is part of the system's environment~\cite{klößner2026intelligentsupporthumanoversight}. Related work models a cooperative human planner to interpret the user’s actions and instructions~\cite{zhi2024pragmatic}. In the \Planner{} we evaluate, the roles are reversed: the human is the listener, and the system plans on the user completing what it omits~\cite{das2026planning}.

This reversal sits uneasily with how human–AI interaction is usually framed. General guidelines ask systems to present information relevant to the user's task and to behave consistently with their social expectations~\cite{amershi2019guidelines}. For conversational systems, pragmatics offers guidance on how information should be communicated. Gricean maxims have been operationalised for referring-expression generation~\cite{dale1995computational}, reinterpreted as design considerations for chatbots and LLM interfaces~\cite{setlur2022converse,miehling_language_2024}, and proposed as a basis for collaborative agents~\cite{saad2025gricean}. Related accounts of cooperative systems argue that the system should reason over a model of its human partner, so that cooperation is designed in rather than emergent~\cite{gmytrasiewicz2001rational,collins_building_2024,miehling_language_2024,saad2025gricean}. For pragmatic assistance, cooperation presupposes that the partner reasons back. Yet a recent systematic review of 51 studies on pragmatics in HCI~\cite{wang2026pragmatics} finds the literature concentrated on evaluating the pragmatic capability of language models, chatbots, and robots.  The human side of the exchange, the listener's contribution, is under-examined. We ask whether people engage in the pragmatic comprehension such systems expect of them.

\subsection{Do People Reason Pragmatically About Systems?}\label{sec:pragmaticsystems}
The answer depends on what people believe their partner to be. And that has changed. A decade ago, human listeners adapted themselves to their computer partner, resolving ambiguous instructions as the speaker would have meant them, at a measurable cognitive cost of doing so~\cite{duran2011listeners}. AI has since profoundly transformed assistive systems, and with them people's perception of them as communicative partners. They now expect a competent machine to have adapted to them, and interpret its message as designed for them, the more readily the more capable it appears~\cite{loy2023perspective}. This is the type of listener a pragmatic assistant assumes, one who expects messages targeted to them and completes them from prior knowledge. But expectations about an AI's competence seem insufficient. Told that an identical ambiguous message came from ChatGPT rather than from an adult, listeners interpreted it less pragmatically, while rating ChatGPT's pragmatic ability highly: capable, but not cooperative~\cite{mayn2024capable}.

This raises a concern. A pragmatic assistant relies on users to complete what it leaves unspecified from their own knowledge~\cite{das2026planning}. Yet a user might instead read ambiguity or omission as evidence that the system has overlooked relevant information, giving them reason to trust its competence less. Rather than using their knowledge to recover what the system meant, they may disregard the message altogether. We therefore examine whether people use their prior knowledge to interpret such messages and act on them in a dynamic, time-critical task.

\section{Method}\label{sec:method}

We examine whether people use prior knowledge to understand an AI assistant’s pragmatic messages and act on them as a situation changes.
We study this in DroneWorld, a dynamic monitoring task in which participants detect critical sensor values on four drones, supported by a voice assistant. Between participants, we vary the type of assistant that supports them: a \Baseline{} specifying every problem in full, and a \Planner{} that adapts what it says to the listener's knowledge. Within participants, we manipulate their prior awareness across trials by providing briefings with different levels of detail before each scene. We analyse participants' clicks, belief probes, and ratings with mixed-effects models to examine both participants' interpretation of alerts with different specificity and their performance across the unfolding task.

\subsection{Task and Interface}\label{sec:method_task}
\begin{figure}
    \centering
    \includegraphics[width=\linewidth]{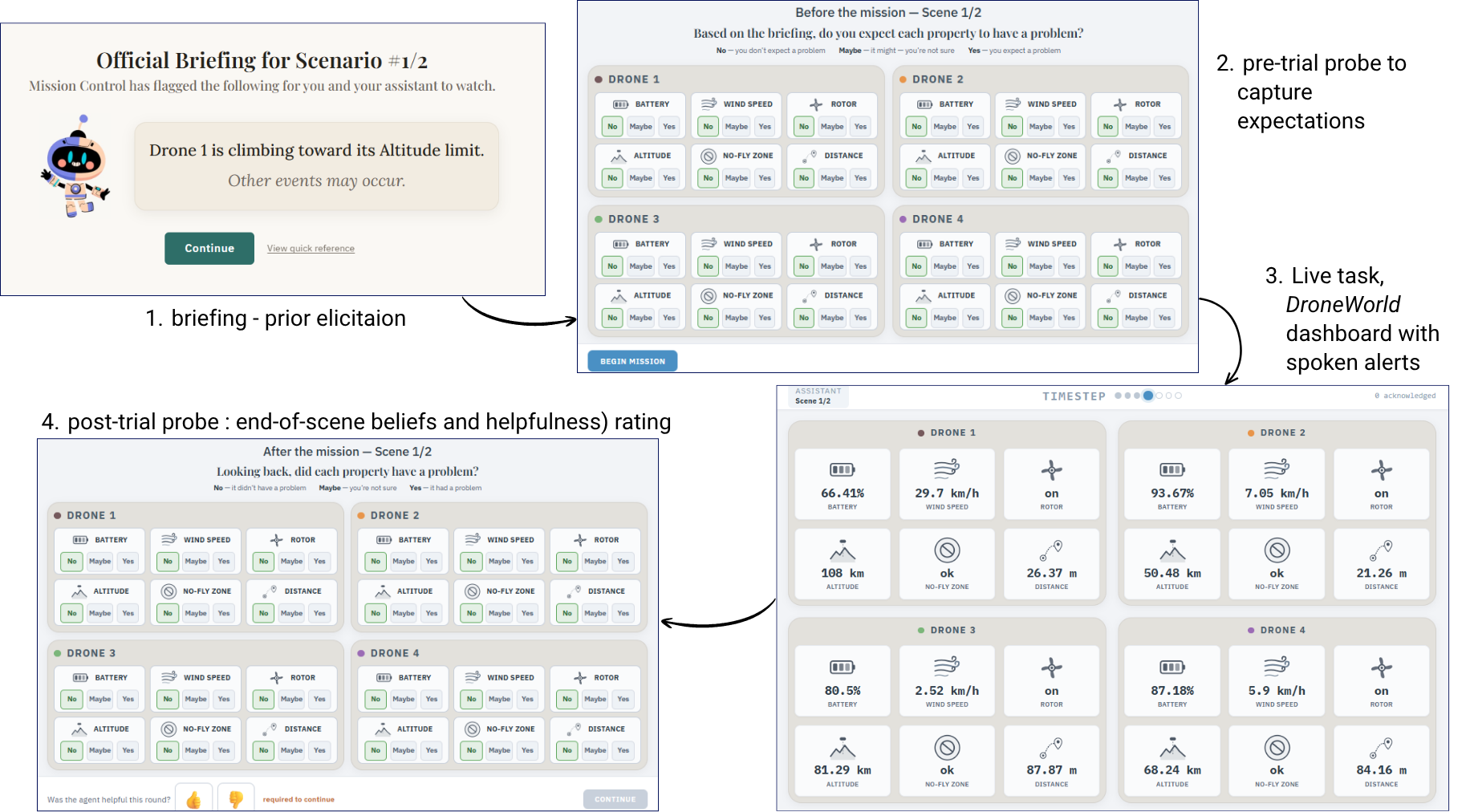}
    \caption{Every trial consists of 4 steps. First, a briefing about the forthcoming scene to set the participant's prior knowledge (awareness). Next, to capture their beliefs pre-trial, a probe marks each of the 24 cells as No (default), Maybe, or Yes for their expectations of criticality. Third, the task scene itself, with an orientation frame (1.5\,s) followed by seven live frames of 2.25\,s each in which alerts play, and participants click the cells they judge critical, ending lastly with a closing (1.5\,s) frame. Clicks are timestamped. A post-trial beliefs probe identical to the first to capture beliefs after the scene, ending finally with a required helpfulness rating of their assistant.}
    \label{fig:m1_interface}
\end{figure}

We develop \textit{DroneWorld}, an environment for studying dynamic reference resolution in the context of a drone monitoring task. Participants are asked to continuously monitor the state of four delivery drones and detect and click on critical sensor values.
The monitoring interface is laid out as a grid of $24$ continuously updating numbers denoting  battery, altitude, wind speed, rotor status, distance, and no-fly-zone status for each of the four drones (Fig.~\ref{fig:m1_interface}). Numbers change at 2.25\,s intervals, and a value turns critical when it crosses a pre-defined threshold. Detection is supported by a voice assistant announcing critical values through natural language messages or auditory alerts.

A trial consists of four steps (see Fig.~\ref{fig:m1_interface}). A participant is first briefed about upcoming critical situations (1). We then validate their situation awareness in a pre-trial probe (2).
Participants are then shown the monitoring task which runs for 17.25\,s during which 3--4 critical situations can occur (3). Finally, participants' situation awareness is registered again in a post-trial probe (4).

We designed the interface and task to allow us to isolate the cognitive process under study: the grid is visually complex but carries no visual salience cues (e.g., colour, highlighting), so a critical event is legible only through two channels: if the raw numerical value had been read or an assistive alert understood. This eliminates the confound of visual attention, making alert comprehension the mechanism by which a reference is resolved. Alerts are spoken, because the manipulation of specificity (regulating how much is said) is expressible distinctly in language. Furthermore, the temporal, fleeting nature of auditory information makes it well-suited for conveying a sequence of alerts that must be processed without visual search \cite{welch1980immediate, wickens2008multiple}.

\subsection{Assistive Agents}\label{sec:method_agents}
The two assistants both observe the same world, share the same channel with finite capacity, thereby being equally incapable of alerting of every critical event, and so, in every scene, leave some problems unspoken. Both share the same vocabulary and are truthful. They depart from each other in \textit{how} and \textit{what} they decide to spend the capacity on.

\textbf{The \BaselineCap{}} is a literal reporter, resembling the communication policy of most traditional alerting systems: it communicates every critical event fully and unambiguously~\cite{isa18_2_2016,eemua191} (e.g., `drone 3, battery low'), in the order it arose until its channel capacity is exhausted, and then falls back to a non-specific tone (e.g., `beep'). It makes no assumption about the listener's knowledge nor of the world. Thus, a well-briefed listener and an uninformed one are supplied with the same alerts in the same order.

\textbf{The \PlannerCap{} (\PP{})} is the d-RSA model of \citet{das2026planning}, described in Sec.~\ref{sec:background}, which we use unchanged. It decides jointly which problems to raise, when, and how explicitly to describe each, by reasoning about what the listener already knows. To a participant who has reason to expect drone 3's battery to fail, the shorter `battery low' is enough, because their prior knowledge completes the reference. Where a participant has no such expectation, the planner names the drone as well, which is what Alice and Bob hear in Fig.~\ref{fig:1}. A fully specified alert takes 1.5 to 3 times as long to speak as an underspecified one, so every reference a participant resolves unaided leaves the planner capacity for a further problem that the \Baseline{} may never reach.

\subsection{Experimental Design}\label{sec:method_design}
We adopt a mixed design. The \emph{assistant type} is a between-participants factor, so each participant worked with either the \Planner{} or the \Baseline{}, avoiding order effects. \emph{Awareness} varies within participants and is manipulated through the briefing shown before each scene, which covered either most (high, 75\%) or only a few (low, 30\%) of that scene's upcoming critical events.
Briefings were worded identically for both assistant groups.

Participants completed 20 scenes in randomised order, ten at each awareness level. Awareness was fixed per scene, so that every scene could be compared across both assistants. We constructed the scenes to cover all 24 grid properties uniformly and to vary the onsets of critical events. A list of scenes with number, type, and order of critical situations is provided in the supplementary material.

\subsection{Participants}
\label{sec:participants}
We recruited $120$ participants through Prolific (desktop, working sound); after preregistered exclusions for data quality, $114$ remained ($M_{Age} = 32.2 \pm 10.2$; \textit{range:} 18--66), 58 from the \Planner{} group and 56 from \Baseline{}. $59$ identified as men, $54$ as women, $1$ non-binary; between them 65.8\% were native English speakers, with the remainder reporting 21 other first languages with fluent English. Participants were from 25 distinct countries, with the largest groups from South Africa, the UK, and the USA. Sessions ran about $M_{Dur} = 40.4$ min ($IQR$ 34.3--49.0; \textit{range:} 20.5--77.5) long, and participants were paid \pounds$8$.

\subsection{Procedure}
\label{sec:procedure}
We ran the study as a game, casting each person as a drone supervisor on a challenging mission, with an AI assistant at their side, but participants were never shown any scores or given any feedback on their detection performance. Using such a cover-story language game to study pragmatic reference helps human listeners conceptualise their partner's role in the human-AI team, and more notably motivates engagement across the length of the study \cite{sikos2021reevaluating}. Participants watched a training video explaining the world and the task, and were then \textit{introduced} to their AI assistant, with both groups receiving a cover story of their agents (see supplementary material). No participant knew a second assistant existed. After familiarising themselves through two practice trials and short comprehension questions, participants worked through $20$ task trials, as explained above. After each trial, they answered one question regarding their perceived helpfulness of the assistant.
A session concluded with a structured ten-item questionnaire on the assistant's communication and the demands of the task, three free-text questions about their experience, and demographics (screenshots in supplementary material). The study was approved by the authors' university ethics board and was preregistered (available at \url{https://osf.io/v79s6/?view_only=5ab6c2fbf9c747c497df9c31af90eeee}).

\subsection{Measures and Analysis}\label{sec:method_analysis}
Our analysis is organised around three questions. Whether participants could act on the alerts, whether the pragmatic comprehension that the planner assumes actually occurred, and how participants experienced the interaction.

The first set of measures is behavioural and captures whether a participant caught the problems and how quickly. \textbf{Detection} is derived from the clicks. A cell counts as caught when it was clicked at or after its critical onset, and a click on a cell that has not yet turned critical counts as a false alarm. \textbf{Speed} or latency is measured as the time from a problem's onset to the first correct click on it. Because the two assistants alert participants of different subsets of problems, we account for events that were never caught. \textbf{Beliefs} are derived from the pre- and post-trial probes and scored as the squared error against the scene's ground truth, averaged over critical cells. Clicks and post-trial beliefs measure the same underlying construct, a participant's representation of the situation, but at two different moments. A click captures a belief as it is acted on, whereas the probe adds a memory step, and in practice the two often disagree.

The second set concerns the mechanism, that is, what participants do with an underspecified alert and how the interpretive work divides between the listener and the message. Every message has a referent set, the property cells it could plausibly denote. A fully specified alert has one, while an ambiguous alert has several, so that `battery low' can refer to any critical battery, for instance those of drone 2 and drone 3. We score an alert as resolved when a participant clicks a cell from that set at or after onset, and record both \textbf{resolution} and \textbf{latency}. On the speaker's side we log what each assistant said about each problem, which lets us hold the same problem fixed across both groups and compare their coverage against participants' responses.

For the third question, we collect data on participants' \textbf{subjective experience} of the interaction from per-scene helpfulness ratings and the post-study questionnaire and free-text prompts.

\paragraph{Analysis} We use generalised linear mixed models (GLMM)\footnote{Fitted in R (lme4) as \texttt{outcome \textasciitilde{} assistant * awareness + complexity + position + (1 | participant) + (1 | scene)}: logistic mixed models (\texttt{glmer}) for detection and speed, a linear mixed model (\texttt{lmer}) for belief error; full specifications are in the supplement.} to model each quantitative measure  as a function of assistant, awareness, their interaction, scene complexity, and trial position, using a link function appropriate to the variable. All models included crossed random intercepts for participant and scene, and predictors were sum-coded. We read the assistant's benefit primarily from the interaction term, because it shows how far awareness continues to shape performance within each group. Effects are reported across participants and the full model formulae, coding, and further analyses are detailed in the supplement.

\section{Findings}\label{sec:result}

Our $114$ participants worked through $2218$ analysed trials, in which $7763$ critical problems arose. Here, we report observations from the user study in four parts: whether the \Planner{} helped, how participants understood its underspecified alerts, how it allocated its speaking budget, and how participants experienced the assistance.

\subsection{Assistance: Detection, Speed and Beliefs}\label{sec:result1}
We first examine whether the \Planner{} (henceforth, \PP{}) improved assistive outcomes. Participants who worked with the \Planner{} (\PP{}) acted correctly on $3329/3879 = 85.8\%$ of critical problems, against $3048/3884 = 78.5\%$ under the \Baseline{}, a gain of $7.3$ percentage points (pp). The adjusted mixed model described in Sec.~\ref{sec:method_analysis} confirmed this difference, $z = 2.95$, $p = .0032$.\footnote{All $p$-values two-sided. Logistic mixed models report $z$, $t$ from Gaussian models and participant-level comparisons. Full estimates and CIs in supplement.} A signal-detection analysis shows this was sharper discernment rather than response bias: \PP{} participants distinguished real problems from spurious ones more reliably ($z = 3.63$, $p < .001$) and did not simply click more indiscriminately (\emph{n.s.}), i.e., no reliable change in response criterion ($z = 1.85$, $p = .064$).

\begin{figure}[b]
    \centering
    \includegraphics[width=\linewidth]{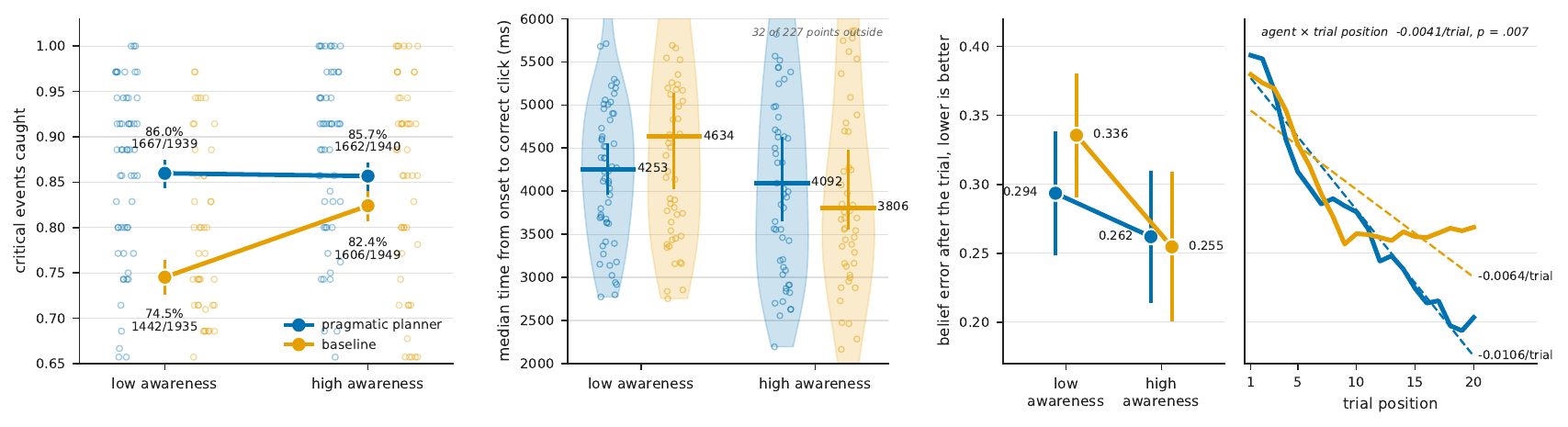}
    \caption{Did it help? Detection, speed, and belief error, by prior awareness and assistant. \PP{}'s advantage is confined to low awareness throughout. (a) share of critical events caught, one point per participant; (b) each participant's median time from a problem's onset to the correct click, the bar marking the median across participants; (c) belief error at the end of the trial, by awareness; (d) the same measure across trial position, falling more steeply under \PP{}.}
    \label{fig:r1_performance}
\end{figure}

The benefit, however, was not uniform. Fig.~\ref{fig:r1_performance}a shows it is concentrated where participants had the least prior knowledge (low awareness = LA) to draw on. For trials where participants arrived already aware (high awareness = HA) of most problems, participants detected $1662/1940 = 85.7\%$ of problems with \PP{} and $1606/1949 = 82.4\%$ with the \Baseline{}, a $3.3$\,pp gap, which was not significantly different ($z = 1.38$, $p = .168$). Whereas under LA, participants detected $1442/1935 = 74.5\%$ with the \Baseline{}, a greater $1667/1939 = 86.0\%$ with \PP{}, i.e., the \PP{} produced a significant gain of $11.5$\,pp, $z = 4.22$, $p < .001$, with a reliable interaction ($z = -4.35$, $p < .001$). Read from the direction of awareness, the finding is surprising: HA was worth $7.9$\,pp of increased detection to a participant working with the \Baseline{} and was not significant (\emph{n.s.}) for one working with the \PP{} ($p = .621$). One might expect a cooperative assistant and prior knowledge to add together; they did not. Instead, with the \PP{} assistant, prior awareness ceased to predict how much a participant caught.

The same pattern held for speed: participants under \PP{} caught problems sooner (a higher catch rate per frame) ($z = 2.35$, $p = .019$), with gains accumulating only under LA (LA $p = .0005$; HA $p = .32$; interaction $z = -4.20$, $p < .001$). Median latencies (Fig.~\ref{fig:r1_performance}b) were on average 548\,ms shorter under LA, $t(86.8) = -2.39$, $p = .019$, Cohen's $d = -0.45$, but only 47\,ms shorter under HA ($p=.874$).

We expected \PP{} to leave participants with more accurate post-trial beliefs. Overall, it did not, but neither did it leave them worse: end-of-scene belief error differed by $-0.020$ (\PP{} minus \Baseline{}), 95\% CI $[-0.089, 0.048]$, $t(115.3) = -0.60$, $p = .552$. Read by awareness, though, the same conditional shape appears here as in detection and speed: prior awareness predicted post-trial belief accuracy more strongly under the \Baseline{} and much less so under \PP{} (interaction $= +0.058$, 95\% CI $[0.001, 0.114]$, $t(20.5) = 2.12$, $p = .046$); averaging over awareness is what
flattens the overall comparison (Fig.~\ref{fig:r1_performance}c). Belief error also declined more steeply across the session under \PP{}, $t(2089.5) = -2.68$, $p = .0073$; see Fig.~\ref{fig:r1_performance}d.

A pragmatic assistant thus helped participants in proportion to how low their prior awareness was. It helped most when a person's awareness was the least: they caught more problems, caught them sooner, and improved more over time; while those who began well-informed were left no worse off.

\subsection{Pragmatic Comprehension: Resolving Underspecification}\label{sec:result2}
\begin{figure}[b]
    \centering
    \includegraphics[width=\linewidth]{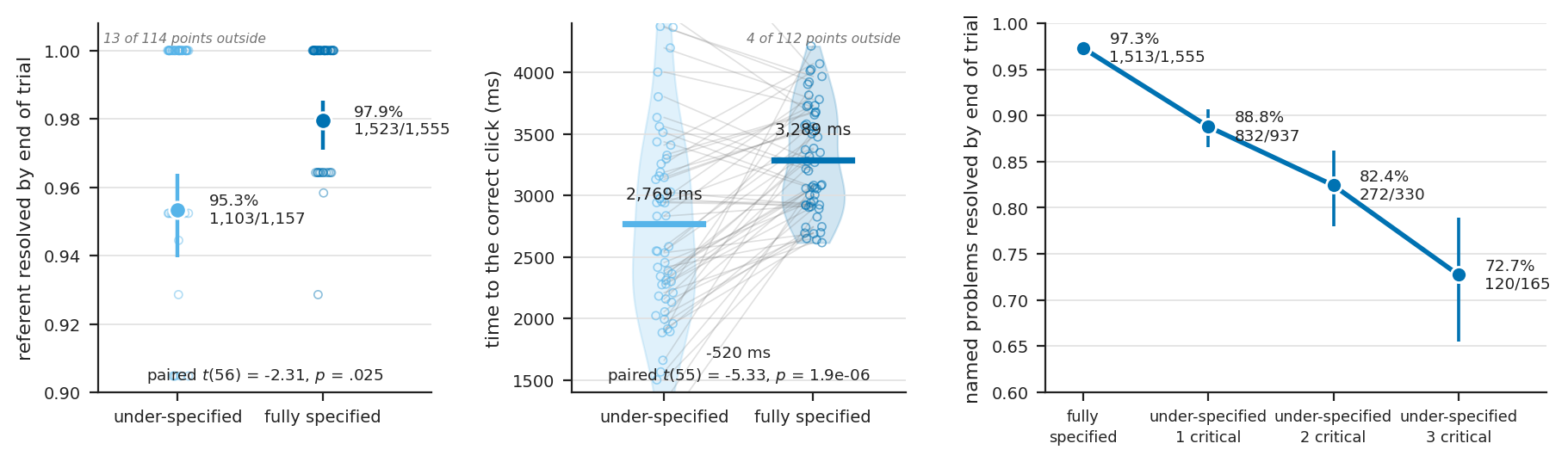}
    \caption{People resolved an underspecified alert almost as often as one that named its exact referent, and sooner (\PP{} only; \Baseline{} never underspecifies). (a) share resolved by end of trial, one point per participant, with the fully specified rate for comparison; (b) time to the correct click; (c) how underspecified alerts can allude to several problems at once, and how many of them people catch.}
    \label{fig:entropy_alerts}
\end{figure}

The planner's benefit comes from two halves of the same exchange working together; this section takes the listener's perspective: what a participant did with an alert that named less than the whole problem.
As Sec.~\ref{sec:method_agents} describes, \PP{}'s communication strategy makes a bet, a probabilistic one, with every underspecified message: trusting the listener's own knowledge will close the reference. Listeners repaid this trust.

An underspecified alert names a problem without its exact location, licensing several grid cells; a listener reasoning from the message alone would select a correct referent on 29.0\% of occasions, near chance. After underspecified alerts, participants' first click was correct 92.7\% of the time $(793/855)$, $t(55) = 44.05$, $p < .001$. By the end of the trial, resolution was nearly as accurate as when an alert was fully specified, reaching the intended referent $1103/1157 = 95.3\%$, below $1523/1555 = 97.9\%$; paired $t(56) = -2.31$, $p = .0246$, Fig.~\ref{fig:entropy_alerts}a. They, however, were temporally advantageous. Surprisingly, participants were roughly 520\,ms quicker from the alert onset to the first correct click when an alert was ambiguous than when it was fully specified, $t(55) = -5.33$, $p < .001$, $d = -0.71$ (Fig.~\ref{fig:entropy_alerts}b).

Much of that time was won before the alert had finished. Roughly three alerts in ten were resolved while the alert was still being spoken, at a rate that did not differ by alert type: 26.3\% (216/820) of the underspecified alerts resolved once they had begun, against 29.5\% (422/1430) of fully specified ones, $p = .12$. Among these, participants began clicking more than a second earlier when the alert was underspecified: 1149\,ms after its onset against 2385\,ms for a fully specified alert, $\Delta = 1236$\,ms, $t(49) = -16.24$, $p < .001$, $d = -2.30$. This shows people's willingness to start acting on a partial message rather than waiting for the alert to name the referent for them.

Resolving that referent themselves carried a processing cost, as expected. Counted from an alert's offset rather than its onset, participants reached the referent of an underspecified alert 620\,ms later than a fully specified one, $t(55) = 6.41$, $p < .001$, $d = 0.86$. Fully specified alerts run 2845\,ms on average against an underspecified one's 1697\,ms, so this inference time fits inside the time the longer alert spends naming the referent outright. Despite the inference the listener supplies being costly, people still do not pay for it in time.

Rather, because underspecified messages designate more than one problem at once, a single message could direct attention across several problems, which participants recovered as a set rather than singly. On messages with two possible critical referents, they caught both on 70\% of occasions, recovering 1.65 of two designated referents, and 2.18 of a three-referent alert's, each exceeding the single referent expected were it a fully specified message ($t(164) = 14.30$ and $t(54) = 13.68$, both $p < .001$). Naturally, the precision of reference resolution declined the wider the alert's coverage:  per-referent catch rates fell from 97.3\% at full specification to 88.8\%, 82.4\% and 72.7\% for ambiguous alerts which left one, two, or three referent sets to be resolved (Fig.~\ref{fig:entropy_alerts}c). Even so, that participants continued to work the larger set underspecified alerts pointed at, rather than stopping where a single click would have satisfied the alert, suggests of them performing prior-driven disambiguation.

In other words, listeners resolved the underspecified messages by weighting possible candidates against their situational and prior knowledge. Participants did it almost as consistently as when alerts were fully specified, faster, though at the cost of some interpretive effort, and across a wider set of problems at once.

\subsection{Allocation: Spending a Fixed Channel}\label{sec:result3}
\begin{figure}[t]
    \centering
    \includegraphics[width=\linewidth]{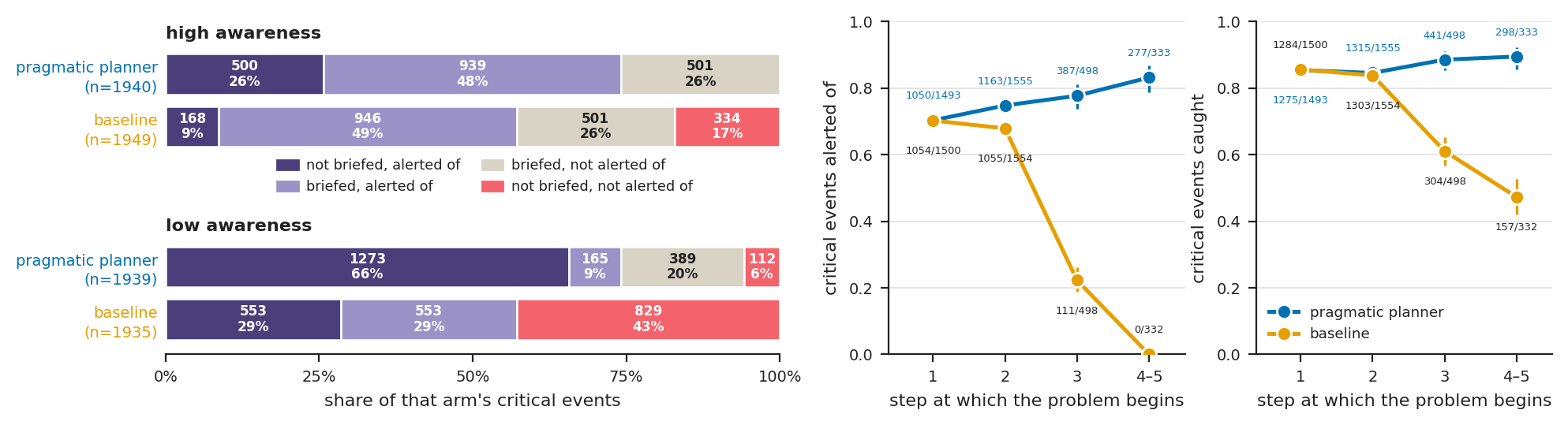}
    \caption{On a comparable channel, \PP{} left far fewer problems for a participant to find unaided. (a) each assistant's alerts, by whether the briefing had already covered the problem; (b, c) whether a problem was alerted and whether it was caught, by the onset step.}
    \label{fig:planning_coverage}
\end{figure}

This section takes the speaker's half of that same exchange: how the planner spent the channel that a listener's own inference set free.
Given the same fixed channel, the two assistants allocated it differently. Across the $20$ scenes, \PP{} uttered 49 messages to the \Baseline{}'s 40, occupying 126 units of speaking against 120 (also counting beep tones, 130 against 132). \PP{} aimed its messages toward problems the briefing had not covered, alerting participants of 94.1\% of these against 38.3\% ($\Delta = 55.8$\,pp). For problems the briefing had already covered, the order reversed: 55.4\% against 75.0\%, Fig.~\ref{fig:planning_coverage}a. The \Baseline{}'s fully specified alerts were hence disproportionately redundant, naming referents participants had already resolved before the alert fired (\Baseline{} 24.5\% against 6.2\% of \PP{}'s fully specified alerts; participant-clustered $t(72.9) = -10.31$, $p < .001$, Cohen's $d = -1.96$). Under HA, we found no advantage for detection (Sec.~\ref{sec:result1}) within assistants; but \PP{} alerted participants of $2.98\times$ more unbriefed problems than the \Baseline{}. People comprehended these unexpected problems accurately, and under HA responded no more slowly overall (47\,ms faster on average, $p = .874$; Sec.~\ref{sec:result1}). Correspondingly, very few critical events reached participants unsupported under \PP{}: only 2.9\% of problems arrived with neither a briefing nor an alert, against 29.9\% under the \Baseline{}.

The \PP{} also reserved capacity for late arrivals (Fig.~\ref{fig:planning_coverage}b). Both alerted at the same rate on the first onset step ($\sim70\%$ each), but the \Baseline{} spends in onset order and runs dry: by the fourth and fifth steps it alerted on none of 332 late problems, while \PP{} alerted on 83.2\% of 333. People caught what it told them. At those late steps, detection held at 89.5\% under \PP{} but only 47.3\% under the \Baseline{}, significantly lower across participants, $\Delta = 42.2$\,pp, $t(100.5) = 8.09$, $p < .001$, Fig.~\ref{fig:planning_coverage}c. Within LA scenes, the contrast sharpens further: a $\Delta = 63.1$\,pp gap, $t(81.2) = 13.03$, $p < .001$.

In summary, while the \Baseline{} spent much of its bandwidth restating the expected, \PP{} spent its channel on the unexpected problems no briefing covered and the ones that arrived late, and participants still succeeded. Pragmatic comprehension held up under conditions that tax it most: alerting of problems one was not prepared for.

\subsection{Experience: Perceived Helpfulness and Cost}\label{sec:result4}
\begin{figure}[h]
    \centering
    \includegraphics[width=0.95\linewidth]{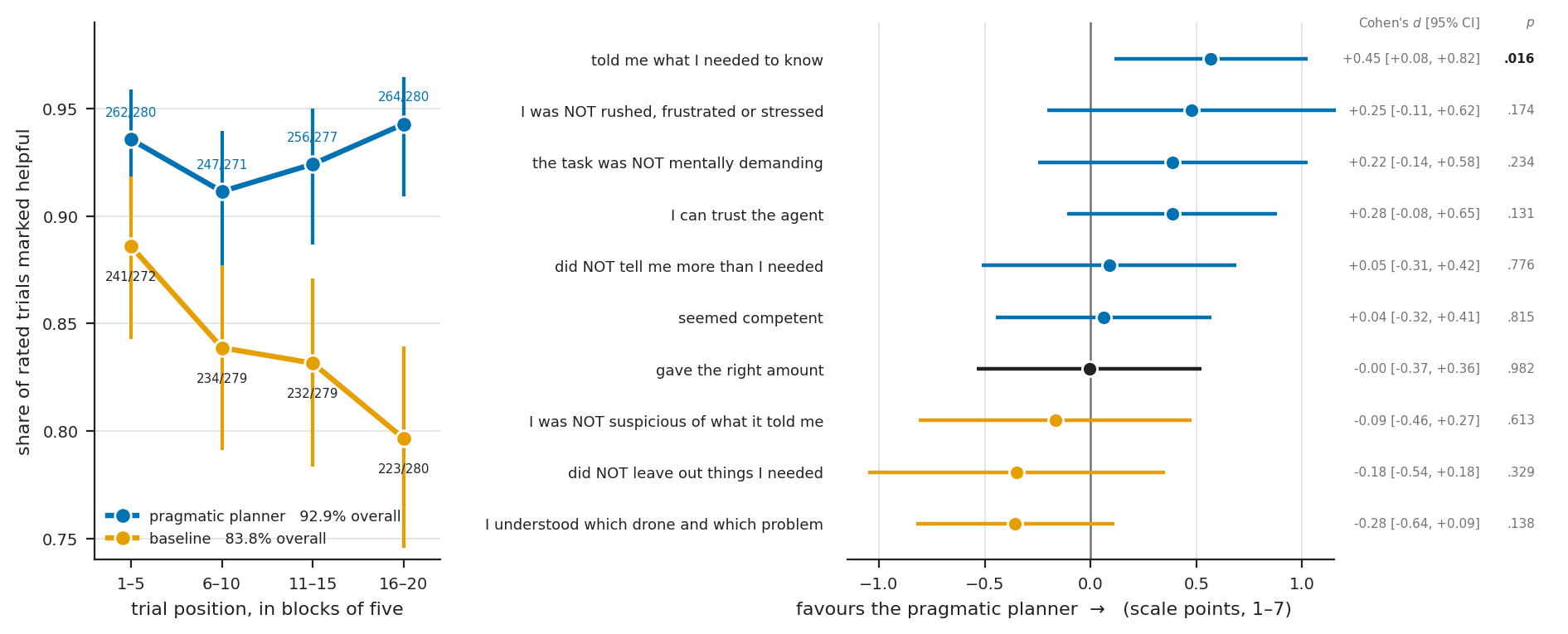}
    \caption{Participants rated \PP{} helpful on more scenes than \Baseline{} users rated theirs, a gap that widened over the session; (a) scenes marked helpful; (b) $10$ items on a scale; five are reverse-scored and reworded, so a $+ve$ value leans toward a better experience with \PP{}.}
    \label{fig:subjective}
\end{figure}

Participants rated the \PP{} helpful (92.9\%) on a higher proportion of scenes than the \Baseline{} (83.8\%), $\Delta = 9.1$\,pp, $z = 2.54$, $p = .011$; per-participant comparison tells the same story: $t(87.5) = 2.73$, $p = .0076$, Cohen's $d = 0.52$. Fig.~\ref{fig:subjective}a shows the two ratings pulled apart as the session went on: people's ratings of \PP{} remained stable across the $20$ scenes, whereas the \Baseline{} declined (interaction $z = 2.80$, $p = .005$; over participants' trial-position slopes: $t(106.7) = 2.29$, $p = .024$, $d = 0.43$).

The felt experience of the task, meanwhile, was not measurably different between the two assistant groups. Of them all, participants agreed more strongly on only one exit-questionnaire item: that the planner told them \textit{what they needed to know} ($t(111.2) = 2.44$, $p = .016$, $d = 0.45$), Fig.~\ref{fig:subjective}b. We strongly suspected that \PP{} leaving its participants to work out references for themselves would cost it, with users finding the task effortful or frustrating. Instead, the task felt much the same under either assistant; participants rated it no more mentally demanding ($p = .23$, $d = -0.22$) and no more frustrating ($p = .17$, $d = -0.25$), and no worse at leaving out things they needed.

To summarise, people found the \Planner{} helpful more often than the \Baseline{} users did theirs, with the conviction widening over the session, and found it no more taxing to work with.

\section{Discussion}\label{sec:discussion}

People performed the comprehension that the pragmatic assistant required of them. They used prior knowledge to resolve underspecified alerts in a dynamic world, allowing the \Planner{} (\PP{}) to direct its finite speaking budget towards problems people's briefings had not covered and problems that arrived late. Across the task, they detected more critical problems and responded sooner than participants working with the \Baseline{}. They also judged the planner helpful more often, with no detected increase in mental demand or frustration. Together, these findings show a form of human–AI cooperation in which people supply meaning where they can and the assistant spends its limited communication on what they still need.

This contributes the human evidence missing from a pragmatic assistance strategy previously evaluated only with simulated listeners~\cite{das2026planning}: what people understood, how they acted, what they reported afterwards, and how they experienced the interaction. Our methodological contribution, \textit{DroneWorld}, helps make the dynamic reference-resolution task observable. We make the environment and accompanying code available at \texttt{[URL UPON ACCEPTANCE]}. Together, our method and findings support a broader contribution to HCI: an assistive system can rely on users' pragmatic comprehension as a practical form of cooperation.

\subsection{People acted more effectively in the moment}\label{sec:discussion1}
People working with \PP{} detected 7.3\,pp (percentage points) more compared to those under the \Baseline{}, and responded on average sooner. The advantage was largest under low awareness, where \PP{} users detected $86.0\%$, against $74.5\%$ with the \Baseline{} (Sec.~\ref{sec:result1}). The allocation results (Sec.~\ref{sec:result3}) show how the \PP{} reached the problems carrying this benefit. The two assistants shared the same channel of fixed capacity but used it dissimilarly. \PP{} alerted participants of 94.1\% of problems their briefings had not covered, compared with only $38.3\%$ under the \Baseline{}. It also retained capacity for $83.2\%$ of late-onset problems, whereas the \Baseline{} alerted participants of none. In a dynamic setting, saying less about a problem the listener could themselves resolve made room to say something about unexpected or upcoming ones they could not. Because selection, timing, and specificity varied together, these results establish the value of the \PP{}'s policy as implemented; they do not isolate underspecification as the cause of the benefit.

Yet this behavioural advantage of detection was not reflected in an overall advantage in the beliefs people reported after each scene. The \PP{}-\Baseline{} difference in post-scene belief error was small and uncertain ($-0.020$ points, $p=.552$). On the other hand, awareness affected belief accuracy, and the interaction ran opposite to the direction we anticipated it would have $(+0.058$ points, $p=.046)$. We predicted that high awareness would especially benefit people working with \PP{}, instead, it mattered much more under the \Baseline{}. Put another way, \PP{}'s apparent advantage was larger under low awareness. In summary, people detected more problems and acted sooner as the scene unfolded, but this improvement did not translate into a reliably more accurate account of the scene afterwards.

P042 (\PP{}) described the separation between immediate assistance and later recollection: \textit{``each one was forgotten as the next came up, but on occasion the agent would pick up an issue that I was not prepared for [...] and wouldn’t have noticed myself.''}

Research on situation awareness distinguishes a person’s understanding of a situation from the task performance that understanding supports~\cite{endsley2017toward, onnasch2014human}. Our measures sampled that relationship at different moments. A click records a belief as the participant acts on it; the post-scene probe asks them to reconstruct the situation after it has passed, with an added memory step. Under time pressure because of the nature of this task, participants may have formed and used local beliefs without maintaining them once a problem had been detected. Alternatively, the \PP{} may have supported timely action without helping them assemble a more complete representation of the scene. The present study cannot distinguish incomplete formation from later loss or retrieval difficulty. P019 (\PP{}) indicated the burdens of such dynamic tasks: \textit{``Honestly the mental task load was high enough that I was not committing things to memory, I struggled to even answer the post-round belief questions [...] All I can say is, the agent did alert me to things I hadn't yet noticed.''}

That distinction matters for the assistance being designed. For events whose opportunity for action expires, helping someone act correctly in the moment is consequential. In settings that require handover, learning, explanation, or later review, however, performance alone cannot establish retained understanding. People may need different communicative support when they need to carry an understanding forward than when they need only act before the moment passes.

\subsection{People can perform pragmatic comprehension in time to act}\label{sec:discussion2}
People resolved referents that \PP{} underspecified, performing pragmatic comprehension while critical problems accumulated and new ones continued to arise (Secs.~\ref{sec:result1},~\ref{sec:result3}). Their first clicks after underspecified alerts provide the clearest evidence of this pragmatic comprehension. Under \PP{}, participants' first clicks substantially exceeded the success expected from chance or even the literal alert alone. A listener relying only on an alert's literal content would select an intended referent on $29.0\%$ of the occasions, participants did so on $92.7\%$, a difference of 63.7\,pp. Similarly, their choices were not explained solely by the numerical display: within an alert’s reference set, a cell previously believed to be critical had $58\%$ higher odds of receiving the first click, even after accounting for its displayed value. By the end of a scene, participants had reached an intended referent after $95.3\%$ of underspecified alerts, 2.6\,pp below fully specified alerts, a small but measurable cost on an otherwise reliable act of pragmatic comprehension.

P057 (\PP{}) described this inference: \textit{``During each mission a particular drone number or problem name was omitted which I think I had to decipher based on prior information at the start of each mission.''} Listeners brought and used information to the interaction that the literal utterance did not contain. Evidence from the first click's latency helps us locate that cost within the timescale of this comprehension. Overall, underspecified alerts saved 1148\,ms of speaking time and, measured from onset, people reached an intended referent 520\,ms sooner. However, participants needed on average 620\,ms longer after an alert ended (offset) to reach the intended referent compared to a fully specified alert, thus capturing the listener's remaining interpretive work. Pragmatic comprehension then took time, but its cost fit within and below the speaking time that underspecification saved. Human–human reference resolution provides a precedent for the pattern observed here: contextual information can guide pragmatic inference early, even when integrating competing information later incurs a processing cost~\cite{richter2020common}.

Earlier evaluation of pragmatic speakers showed that people could benefit from pragmatically chosen references and instructions~\cite{fried2017unified}. Such static (untimed, turn-based) tasks left open whether people could sustain that interpretive work while events developed dynamically and independently of their actions and opportunities to respond expired. Here, they could. Our study places that interpretive work within an interface where events continue to develop, speaking consumes task time, and the communicative partner is an AI assistant. The last condition here also matters, on account of Sec.~\ref{sec:pragmaticsystems}, where we laid out that AI partners do not invariably invite pragmatic interpretation. Our participants nevertheless performed pragmatic comprehension without a corresponding advantage for \PP{} in perceived competence or trust: neither judgement reliably distinguished the assistants. Instead, these ratings correlated more closely with participants' broader views of AI communication tools. Those who regarded AI tools as generally good communicators rated their own assistant as more competent ($5.71$ vs $4.95$, $d=0.56$, $p=.0058$). Unlike \citet{mayn2024capable}, however, the interaction was framed as a collaborative task: a third party \textit{Mission Control} supplied the participant and assistant with the same briefing, establishing task-specific common ground from which they worked as partners. Participants were also told only that the AI assistant would help them. Our results therefore do not overturn prior work, but identify a setting in which an AI assistant need not be regarded as unusually competent or trustworthy for people to engage in pragmatic comprehension with it.

\subsection{People can value assistance that leaves them a part to play}\label{sec:disc_cost}
People recognised useful assistance even when it left them in an interpretive role with referents to work out. The \Planner{}'s helpfulness ratings remained stable over the session while the \Baseline{}'s declined; they agreed more strongly that it had told them what they needed to know; they found the task no more demanding and no more frustrating, if anything the reverse (Sec.~\ref{sec:result4}); and their picture of the scene afterwards was no worse (Sec.~\ref{sec:result1}).

We had expected the additional interpretation to feel burdensome. Interfaces that make people think harder are rated worse for it~\cite{buccinca2021trust}; people preferred unambiguous instructions even when the shorter, interactive alternative was faster~\cite{mitev2018using}; communicative effort compounds between agent and person~\cite{pena2026help}; and a speech agent that adapts its language case by case has been judged a worse partner than a rigid one~\cite{edwards2024comparing}. Although these were different interventions with different costs, and ours measured a subset of these, on what we measured, no cost appeared. The possibility that follows is a hypothesis, and the one we think most worth testing. Complacency comes from an assistant that does the person's work~\cite{onnasch2014human}; distrust or a loss of autonomy from one that does the wrong work, or less than it promised~\cite{faas2024give}. An assistant that leaves the right work undone, the part the person can do from what they know, may occupy neither failure mode, as here, where people could make productive use of underspecification to stay involved.

For P057 (\PP{}), having a part to play was part of the appeal: \textit{``I enjoyed providing answers myself, it helps me feel like I have accomplished something on my own''}. Yet participation was not welcome in every form. P091 (\PP{}) explained: \textit{``If it knows there is a problem with a particular drone, then being specific is more useful and helps to deal with the issue faster. I feel that only partly providing an answer slowed me down.''} These reflections describe different experiences of the same policy, rather than a shared preference for having work left to do. Their value for design is the distinction between an inference someone finds worthwhile and an omission they experience as an impediment.

\subsection{Limitations and Future Work}\label{sec:disc_limit}
The dissociation between action and post-scene belief is one clear direction for a following study. The present design cannot yet tell why participants created local belief needed for an immediate click, but then not for the post-trial belief probe screen. They either later forgot it, or retained the relevant beliefs but the post-scene probe failed to retrieve them. A design that probes the same event immediately after the click, and again at scene end, would separate them. If the advantage is only ever a correct reflex, it would already be absent from immediate probing. This would show when pragmatic assistance can support learning, handover, and later review as well as action in the moment.

A second study should relax an assumption ours had to deliberately fix. Briefings gave experimental control over what participants had been told, and the planner's estimate was accurate by construction, because awareness was set by the briefing. A deployed system has no equivalent guarantee: it must infer what a person noticed, retained or already knew from behavioural signals, an estimate that will sometimes be wrong. So it should vary the accuracy of this estimate and the confidence of that inference independently, to locate where underspecification stays recoverable and where it does not.

The task and design choice opens a specific extension. Critical changes remained numerically visible but received no specific visual salience at onset, so that spoken reference resolution could be observed without a competing visual cue. Restoring such cues would test whether prior-guided interpretation still contributes when the interface can direct attention visually. Broader scene sets, other tasks in which communication itself consumes consequential time, and longer interactions would reveal how the policy fares across events or domains to see whether people get better, worse, or just tired. Clinics and safety-critical settings motivate and make this question worth asking in the first place, they remain settings for future validation. What remains is more defined, that our findings are useful where an assistant can make a defensible guess about what its user knows, where communication or alerting costs real time or attention, and where the world keeps moving, new information can arriving or expiring while communication is still underway.

\section{Conclusion}
In this work, we asked whether people complete an assistant's underspecified messages with their prior knowledge or beliefs, while the world evolving gives communication and action no time to spare. Across the $114$ participants we present results from, they did. Message by message, the \Planner{}'s underspecified alerts were resolved almost as reliably as its fully specified ones, and sooner from onset. Because that comprehension held, the time it freed could be spent where the \Baseline{} could not afford to, on the unbriefed and late-arriving problems each participant's briefing did not cover. The \Planner{}'s users caught more of them, caught them sooner, and judged the assistance helpful more often. That repositions pragmatic comprehension under these conditions: not just a capability proven against a simulated listener, but a demonstrated form of cooperation with the people who actually have to act on what a message leaves unsaid. When the world will not wait, effective assistance is not a matter of saying more. It is a matter of judging what a person can supply, and spending what that judgement saves on whatever comes next.

\appendix
\newcommand{\bhMin}{.148}
\newcommand{\binAmb}{.101}
\newcommand{\binSpec}{8.6\times10^{-4}}
\newcommand{\briefHA}{26 of 35}
\newcommand{\briefHApct}{74.3\%}
\newcommand{\briefLA}{10 of 35}
\newcommand{\briefLApct}{28.6\%}
\newcommand{\chance}{$29.0$}
\newcommand{\clipS}{2,845}
\newcommand{\clipSrange}{2,640--3,024}
\newcommand{\clipU}{1,697}
\newcommand{\clipUrange}{1,488--1,968}
\newcommand{\coxAmb}{.282}
\newcommand{\coxSpec}{.0012}
\newcommand{\emmeansVersion}{1.11.2}
\newcommand{\ftN}{117}
\newcommand{\ftResp}{351}
\newcommand{\hThreeN}{607}
\newcommand{\hThreeRet}{526}
\newcommand{\intPone}{.977}
\newcommand{\intPtwo}{.046}
\newcommand{\latLAn}{58 and 56}
\newcommand{\licCorrect}{1.26}
\newcommand{\licMean}{4.36}
\newcommand{\lmeVersion}{1.1.35.5}
\newcommand{\lmerTestVersion}{3.1.3}
\newcommand{\nAnalysed}{2,218}
\newcommand{\nAnswered}{855}
\newcommand{\nCells}{53,232}
\newcommand{\nCompBL}{58}
\newcommand{\nCompPP}{59}
\newcommand{\nCompleted}{117}
\newcommand{\nEvents}{7,763}
\newcommand{\nNoTrial}{3}
\newcommand{\nNoTrialBL}{2}
\newcommand{\nNoTrialPP}{1}
\newcommand{\nRemoved}{122}
\newcommand{\nRemovedBL}{50}
\newcommand{\nRemovedPP}{72}
\newcommand{\nRetTrials}{2,014}
\newcommand{\nRulePart}{110}
\newcommand{\nTrials}{2,340}
\newcommand{\orPrior}{1.58}
\newcommand{\orPriorYesNo}{2.48}
\newcommand{\orPriorYesNoCI}{[1.85, 3.34]}
\newcommand{\provH}{5}
\newcommand{\provHlist}{HA\_nc4\_1, LA\_nc3\_6, LA\_nc4\_3, LA\_nc4\_7, LA\_nc4\_9}
\newcommand{\provM}{8}
\newcommand{\provR}{7}
\newcommand{\rVersion}{4.4.1}
\newcommand{\redD}{-1.96}
\newcommand{\retHA}{88.0\% (978/1,111)}
\newcommand{\retLA}{93.6\% (1,036/1,107)}
\newcommand{\sensHazEveryP}{.096}
\newcommand{\sensRetBel}{.576}
\newcommand{\sensRetDet}{.0030}
\newcommand{\sensRetHaz}{.016}
\newcommand{\sensRuleBel}{.645}
\newcommand{\sensRuleDet}{7.9\times10^{-4}}
\newcommand{\sensRuleHaz}{.0053}
\newcommand{\unitsBL}{7,186}
\newcommand{\unitsPP}{7,113}

\newcolumntype{L}[1]{>{\raggedright\arraybackslash}p{0.86\dimexpr#1\relax}}%
\setlength{\LTcapwidth}{\linewidth}
\setlength{\tabcolsep}{3.5pt}
\renewcommand{\arraystretch}{1.12}
\renewcommand{\thesection}{S\arabic{section}}
\renewcommand{\thetable}{S\arabic{table}}

\section*{Supplementary Material}

This appendix gives the stimuli and both assistants' alert schedules, each reported analysis with its
status relative to the preregistration, and the fixed effects of the fitted models.

\section{Conventions}

\PP{} is the \Planner{} and the \Baseline{} is the threshold-FIFO reporter. HA and LA are the high- and
low-awareness briefings, which covered \briefHA{} and \briefLA{} critical events across their ten scenes
each. Retention, the share of trials on which participants' pre-trial ratings showed they remembered the
briefing, was \retLA{} after LA briefings and \retHA{} after HA ones. An alert is fully specified when it names a drone and a property (``drone 3, battery low'') and
underspecified when it names only one of them. Differences are \PP{} minus \Baseline{}; on belief error, a Brier
score, a negative value favours \PP{}. All $p$-values are two-sided and brackets give 95\% confidence intervals.
Logistic mixed models report $z$, and Gaussian mixed models report $t$ with Satterthwaite degrees of
freedom. Raw percentages and model estimates are different quantities and are not interchangeable.

\section{Stimuli and the two assistants' schedules}

Each scene runs for seven frames of 2.25\,s after a 1.5\,s orientation frame. An alert occupies the channel
for as many frames as it costs, three for a fully specified alert, two for an underspecified one and one for
a beep, and no alert can begin while another occupies the channel. Both assistants are therefore bound by the
same seven frames per scene. Recorded clips last \clipS\,ms on average for fully specified alerts and
\clipU\,ms for underspecified ones.

\par\medskip
\noindent\begin{minipage}{\linewidth}%
\includegraphics[height=0.39\linewidth]{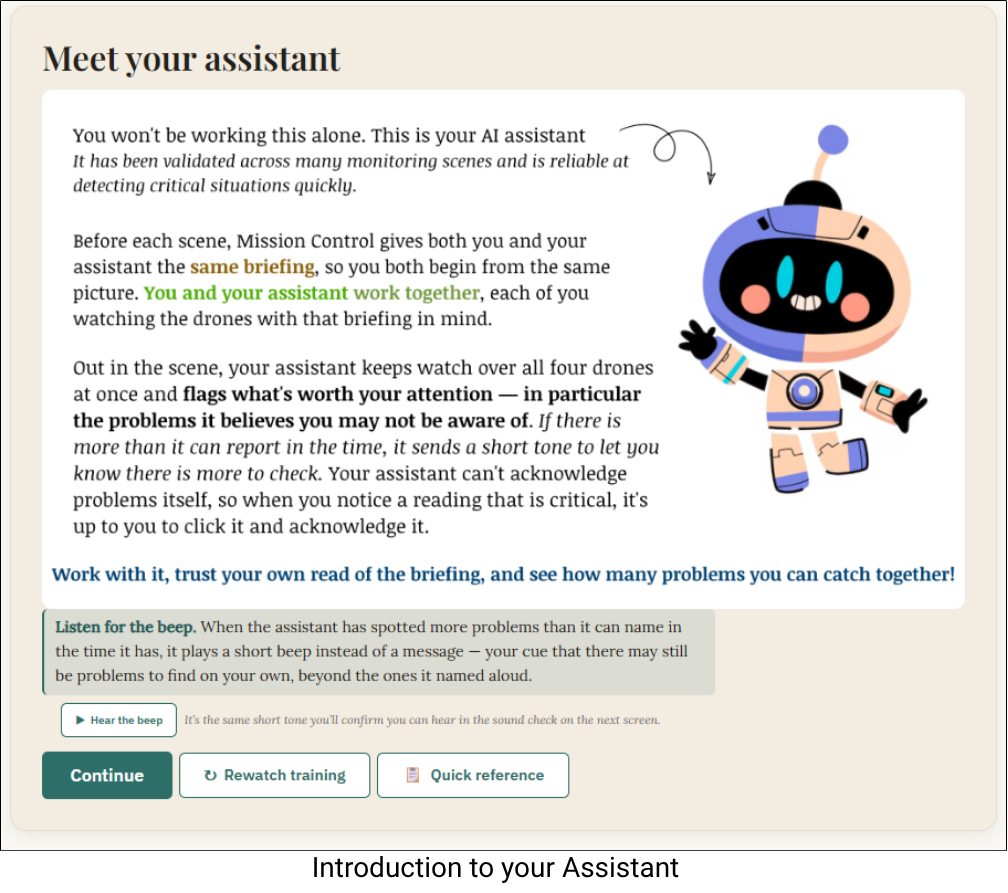}\hfill
\includegraphics[height=0.39\linewidth]{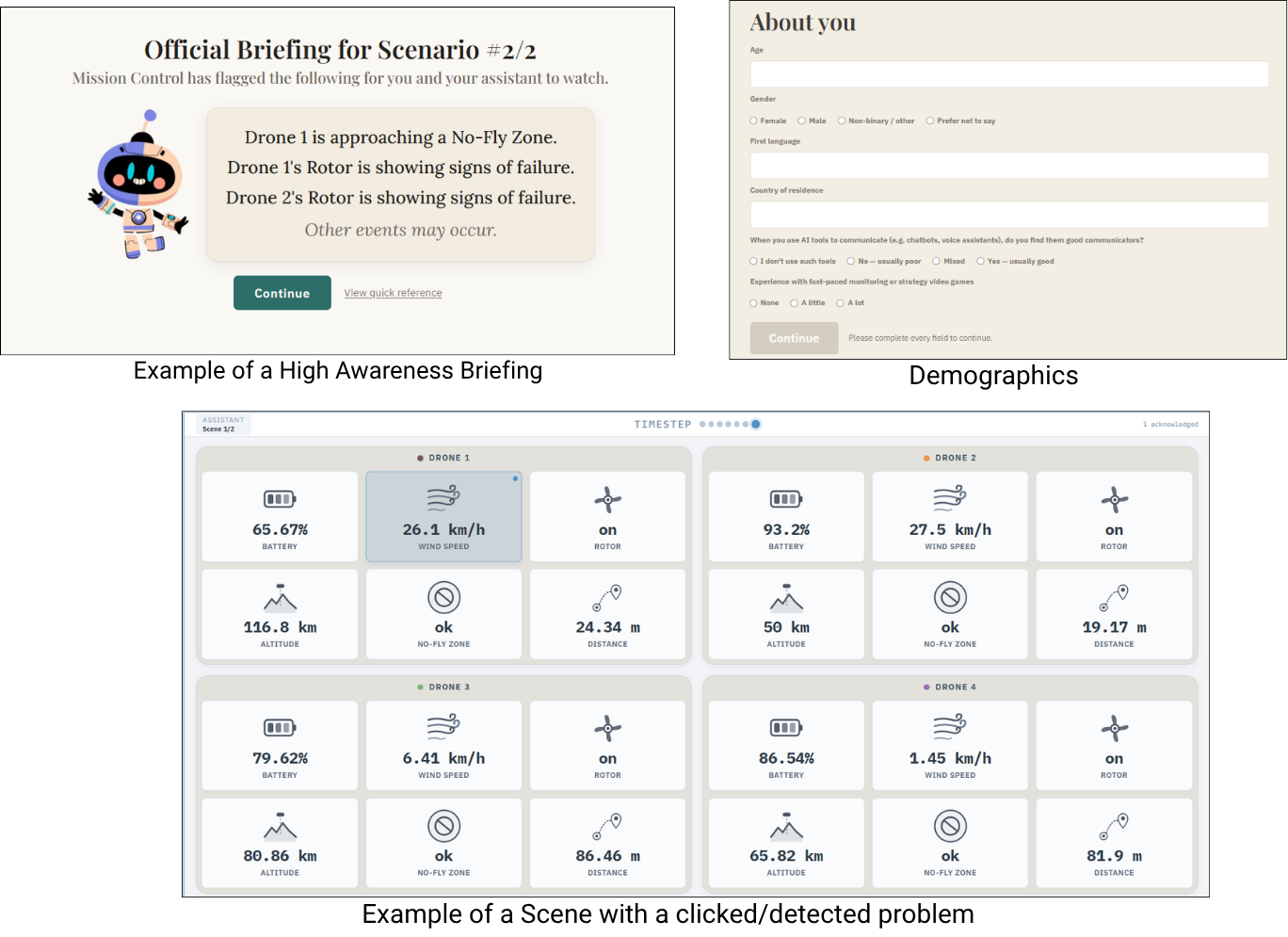}\par\medskip
\includegraphics[width=\linewidth]{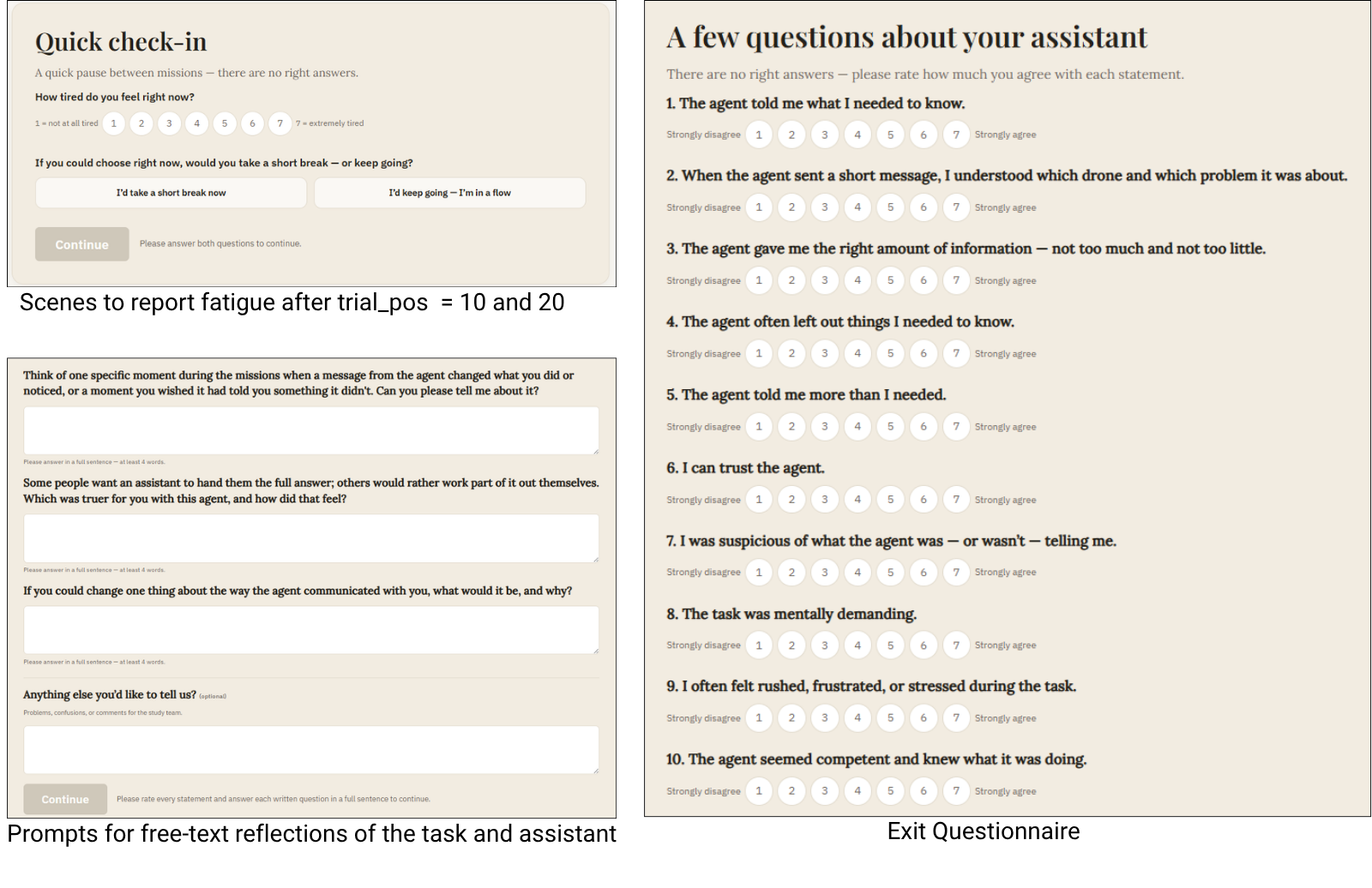}
\end{minipage}\par\medskip

The twenty scenes derive from the drone-supervision scenarios on which d-RSA was evaluated in simulation
\cite{das2026planning}. Ten carry the HA briefing and ten the LA briefing; within each, five scenes have three
critical events and five have four, seventy in all (Table~\ref{tab:scenes_short}). Each scene keeps the same
briefing for every participant, and each participant meets all twenty in an individually randomised order.
The awareness main effect is therefore inseparable from the scenes that carry each level. The agent contrast
is not, because both arms meet every scene, and the agent $\times$ awareness interaction compares that
contrast across the two fixed sets of scenes.

Each assistant's schedule is fixed per scene and identical for every participant in its arm. The \Baseline{}
speaks in onset order, naming each critical event in full while frames remain, and fills a remainder too short
for a full alert with a beep. The planner's schedules originate in d-RSA's output for each scenario. All
stimulus preparation was completed before the main data collection.

\begingroup
\scriptsize
\begin{longtable}{L{1.35cm}L{4.9cm}L{4.95cm}L{4.6cm}}
\caption{The twenty scenes and both assistants' alert schedules.}\label{tab:scenes_short}\\
\toprule
Scene & Critical events (onset frame; * briefed) & \PP{} schedule (start frame, alert) & \BaselineCap{} schedule (start frame, alert) \\
\midrule
\endfirsthead
\multicolumn{4}{@{}l}{\textit{Table~\ref{tab:scenes_short}, continued}}\\
\toprule
Scene & Critical events (onset frame; * briefed) & \PP{} schedule (start frame, alert) & \BaselineCap{} schedule (start frame, alert) \\
\midrule
\endhead
\bottomrule
\endlastfoot
HA\_nc3\_1 & D1 Altitude (2)*; D2 Battery (2)*; D2 No-fly zone (2)* & 1\,--; 2\,Battery; 4\,Altitude; 6\,No-fly zone & 1\,--; 2\,D1 Altitude; 5\,D2 Battery \\
HA\_nc3\_3 & D3 Altitude (1); D4 Wind speed (1)*; D4 Distance (1)* & 1\,D3 Altitude; 4\,D4 Distance; 7\,-- & 1\,D4 Distance; 4\,D4 Wind speed; 7\,beep \\
HA\_nc3\_4 & D1 Rotor (1)*; D2 Rotor (2)*; D2 Battery (3) & 1\,Drone 1; 3\,D2 Battery; 6\,Rotor & 1\,D1 Rotor; 4\,D2 Rotor; 7\,beep \\
HA\_nc3\_6 & D1 Wind speed (2); D3 Rotor (2)*; D4 Altitude (2)* & 1\,--; 2\,Altitude; 4\,D1 Wind speed; 7\,beep & 1\,--; 2\,D1 Wind speed; 5\,D3 Rotor \\
HA\_nc3\_7 & D2 Distance (2)*; D3 No-fly zone (3)*; D4 Wind speed (4) & 1\,--; 2\,Distance; 4\,D4 Wind speed; 7\,beep & 1\,--; 2\,D2 Distance; 5\,D3 No-fly zone \\
HA\_nc4\_1 & D1 Altitude (1)*; D2 Battery (1); D2 Distance (1)*; D4 Wind speed (1)* & 1\,D2 Battery; 4\,Altitude; 6\,Wind speed & 1\,D1 Altitude; 4\,D2 Battery; 7\,beep \\
HA\_nc4\_2 & D1 Battery (1)*; D3 Wind speed (2)*; D3 Distance (3); D4 Altitude (4)* & 1\,Battery; 3\,D3 Distance; 6\,Altitude & 1\,D1 Battery; 4\,D3 Wind speed; 7\,beep \\
HA\_nc4\_3 & D1 Rotor (2)*; D1 No-fly zone (2)*; D2 Rotor (2)*; D3 Battery (2) & 1\,--; 2\,D3 Battery; 5\,D1 No-fly zone & 1\,--; 2\,D1 No-fly zone; 5\,D1 Rotor \\
HA\_nc4\_6 & D1 Battery (1)*; D4 Rotor (2); D4 Altitude (3)*; D4 No-fly zone (4)* & 1\,Battery; 3\,D4 Rotor; 6\,Altitude & 1\,D1 Battery; 4\,D4 Rotor; 7\,beep \\
HA\_nc4\_7 & D1 Altitude (1)*; D2 Battery (1)*; D3 Wind speed (1)*; D4 No-fly zone (1) & 1\,Altitude; 3\,D4 No-fly zone; 6\,Battery & 1\,D1 Altitude; 4\,D2 Battery; 7\,beep \\
LA\_nc3\_1 & D1 Battery (1)*; D2 No-fly zone (2); D2 Distance (3) & 1\,--; 2\,D2 No-fly zone; 5\,D2 Distance & 1\,D1 Battery; 4\,D2 No-fly zone; 7\,beep \\
LA\_nc3\_2 & D1 Altitude (2); D4 No-fly zone (2); D4 Distance (2)* & 1\,--; 2\,D4 No-fly zone; 5\,D1 Altitude & 1\,--; 2\,D1 Altitude; 5\,D4 Distance \\
LA\_nc3\_3 & D1 Battery (1); D1 Wind speed (1)*; D3 Distance (1) & 1\,D1 Battery; 4\,D3 Distance; 7\,-- & 1\,D1 Battery; 4\,D1 Wind speed; 7\,beep \\
LA\_nc3\_5 & D1 Wind speed (1); D1 Altitude (3)*; D2 Wind speed (5) & 1\,D1 Wind speed; 4\,beep; 5\,D2 Wind speed & 1\,D1 Wind speed; 4\,D1 Altitude; 7\,beep \\
LA\_nc3\_6 & D2 Distance (1)*; D3 Rotor (2); D4 Wind speed (4) & 1\,--; 2\,D3 Rotor; 5\,D4 Wind speed & 1\,--; 2\,D2 Distance; 5\,D3 Rotor \\
LA\_nc4\_1 & D1 Altitude (2)*; D2 Wind speed (2); D2 Distance (2); D3 No-fly zone (2) & 1\,--; 2\,D2 Distance; 5\,D2 Wind speed & 1\,--; 2\,D1 Altitude; 5\,D2 Wind speed \\
LA\_nc4\_3 & D2 Rotor (1)*; D3 Rotor (1); D4 Wind speed (3); D4 Distance (3) & 1\,Rotor; 3\,D4 Distance; 6\,Wind speed & 1\,D2 Rotor; 4\,D3 Rotor; 7\,beep \\
LA\_nc4\_4 & D1 Altitude (1); D2 Battery (1)*; D2 Distance (1); D4 Wind speed (1) & 1\,D2 Distance; 4\,D4 Wind speed; 7\,beep & 1\,D1 Altitude; 4\,D2 Battery; 7\,beep \\
LA\_nc4\_7 & D1 No-fly zone (1); D3 Battery (2)*; D3 Distance (2); D4 Battery (3) & 1\,Drone 1; 3\,Battery; 5\,D3 Distance & 1\,D1 No-fly zone; 4\,D3 Battery; 7\,beep \\
LA\_nc4\_9 & D3 Battery (2); D3 Altitude (2)*; D3 Distance (2); D4 Altitude (4) & 1\,Drone 3; 3\,Battery; 5\,D4 Altitude & 1\,--; 2\,D3 Altitude; 5\,D3 Battery \\
\end{longtable}
\vspace{-8pt}{\scriptsize Scene codes begin with the briefing (HA or LA). A fully specified alert (``D2 Battery'') occupies three frames, an underspecified alert (``Battery'', ``Drone 1'') two, and a beep or a scheduled silence (--) one.\par}
\endgroup

\section{Analysis register}

Table~\ref{tab:register_short} lists each analysis the paper reports with its status relative to the
preregistration: a registered test, a registered companion (a model the registration attached to a test),
registered exploratory, or additional. Registered analyses are reported as registered, whatever they
returned. Where the registration specified a one-sided test, the table gives its $p$ beside the two-sided
value the paper reports; no conclusion changes.

Three points connect the registration to the analyses. First, it predicted a larger awareness benefit with
\PP{}; the interaction ran the other way, with \PP{}'s advantage larger under LA. Second, the analyses use every
trial that passes the engagement rule (a trial is removed if the participant made no click, or left a belief
grid untouched), where the registration specified briefing-retained trials and a participant-level exclusion
rule. Refitted on those two populations, detection ($p = \sensRetDet$ and $\sensRuleDet$) and speed
($p = \sensRetHaz$ and $\sensRuleHaz$) remain reliably better with \PP{}, and belief error still shows no
reliable difference ($p = \sensRetBel$ and $\sensRuleBel$). Third, the paper keeps redundant alerts, those
whose referents had already been clicked, in its alert-level analyses, where the registration dropped them;
the registered coding gives the same conclusion.

\begingroup
\scriptsize
\begin{longtable}{L{3.45cm}L{1.55cm}L{3.2cm}L{1.95cm}lL{4.2cm}}
\caption{Each reported analysis and its status relative to the preregistration.}\label{tab:register_short}\\
\toprule
Analysis (unit; model) & Status & Estimate [95\% CI] & Test & $p$ & Registration: prediction; one-sided $p$; outcome \\
\midrule
\endfirsthead
\multicolumn{6}{@{}l}{\textit{Table~\ref{tab:register_short}, continued}}\\
\toprule
Analysis (unit; model) & Status & Estimate [95\% CI] & Test & $p$ & Registration: prediction; one-sided $p$; outcome \\
\midrule
\endhead
\bottomrule
\endlastfoot
\addlinespace[3pt]\multicolumn{6}{@{}l}{\textit{Assistance}} \\
Detection, \PP{} $-$ \Baseline{}\newline\textit{Event; logistic GLMM (M1)} & Registered test & $+0.55$ $[0.18, 0.91]$ (OR 1.73) & $z = 2.95$ & $.0032$ & H1b; \PP{} higher; one-sided $p = .0016$; as registered \\
Detection under low awareness\newline\textit{M1 contrast} & Follow-up & $+0.83$ $[0.44, 1.21]$ (OR 2.28) & $z = 4.22$ & $2.4\times10^{-5}$ & -- \\
Detection under high awareness\newline\textit{M1 contrast} & Follow-up & $+0.27$ $[-0.12, 0.66]$ (OR 1.31) & $z = 1.38$ & $.168$ & -- \\
Detection, agent $\times$ awareness\newline\textit{M1 term} & Registered model term & $-0.55$ $[-0.80, -0.30]$ & $z = -4.35$ & $1.4\times10^{-5}$ & opposite to H2a's prediction \\
Criterion (clicks on non-critical cells)\newline\textit{Cell; probit GLMM} & Registered companion & $+0.14$ $[-0.01, 0.29]$ & $z = 1.85$ & $.064$ & H1b companion \\
Discrimination\newline\textit{Cell; probit GLMM} & Registered companion & $+0.18$ $[0.08, 0.28]$ & $z = 3.63$ & $2.8\times10^{-4}$ & H1b companion \\
Speed, \PP{} $-$ \Baseline{}\newline\textit{Event $\times$ interval; logistic hazard (M2)} & Registered test & $+0.22$ $[0.04, 0.41]$ (HR 1.25) & $z = 2.35$ & $.019$ & H1c; \PP{} faster; one-sided $p = .0093$; as registered \\
Speed under low awareness\newline\textit{M2 contrast} & Follow-up & $+0.35$ $[0.15, 0.54]$ (HR 1.42) & $z = 3.49$ & $4.9\times10^{-4}$ & -- \\
Speed under high awareness\newline\textit{M2 contrast} & Follow-up & $+0.10$ $[-0.10, 0.29]$ (HR 1.10) & $z = 1.00$ & $.320$ & -- \\
Speed, agent $\times$ awareness\newline\textit{M2 term} & Registered model term & $-0.25$ $[-0.36, -0.13]$ & $z = -4.20$ & $2.6\times10^{-5}$ & opposite to H2a's prediction \\
Onset-to-catch latency, low awareness (ms)\newline\textit{Person; Welch $t$ on medians} & Additional & $-548$ $[-1004, -92]$ & $t(86.8) = -2.39$ & $.019$ & -- \\
Belief error, \PP{} $-$ \Baseline{}\newline\textit{Trial; Gaussian LMM (M3)} & Registered test & $-0.020$ $[-0.089, 0.048]$ & $t(115.3) = -0.60$ & $.552$ & H1a; \PP{} lower; one-sided $p = .276$; sign as registered; not reliable \\
Belief error, agent $\times$ awareness\newline\textit{M3 term} & Registered test & $+0.058$ $[0.001, 0.114]$ & $t(20.5) = 2.12$ & $.046$ & H2a; larger awareness benefit with \PP{}; one-sided $p = .977$; opposite \\
Belief error, agent $\times$ retention\newline\textit{Trial (all trials); Gaussian LMM} & Registered test & $-0.015$ $[-0.085, 0.055]$ & -- & $.671$ & H2b; \PP{} gap grows with retention; not reliable \\
Belief error, agent $\times$ trial position\newline\textit{Trial; LMM (M6)} & Additional & $-0.0041$ $[-0.0071, -0.0011]$ & $t(2089.5) = -2.68$ & $.0073$ & -- \\
Detection, agent $\times$ trial position\newline\textit{Event; GLMM (M7)} & Additional & $-0.0047$ $[-0.0264, 0.0170]$ & $z = -0.43$ & $.670$ & -- \\
Decoding hazard: \PP{} underspecified vs fully specified\newline\textit{Alert $\times$ 1\,s bin; logistic hazard (M5)} & Registered companion & $+0.11$ $[-0.02, 0.23]$ & $z = 1.64$ & $.101$ & H1c companion; much the same pace; as expected \\
Decoding hazard: \Baseline{} vs \PP{} fully specified\newline\textit{Alert $\times$ 1\,s bin; logistic hazard (M5)} & Registered companion & $+0.42$ $[0.17, 0.67]$ & $z = 3.33$ & $8.6\times10^{-4}$ & H1c companion; same pace across agents; not as expected \\
Fully specified alerts answered, \PP{} $-$ \Baseline{}\newline\textit{Alert $\times$ 1\,s bin; logistic hazard} & Additional & $-0.46$ $[-0.71, -0.20]$ & $z = -3.45$ & $5.6\times10^{-4}$ & -- \\
\quad adjusting for a briefed referent\newline\textit{Alert $\times$ 1\,s bin; logistic hazard} & Additional & $+0.03$ $[-0.26, 0.32]$ & $z = 0.20$ & $.838$ & -- \\
\addlinespace[3pt]\multicolumn{6}{@{}l}{\textit{Pragmatic comprehension}} \\
Resolved by end of trial, underspecified vs fully specified (\PP{})\newline\textit{Person; paired $t$} & Registered measure, additional coding & $-2.6$ pp $[-4.9, -0.4]$ & $t(56.0) = -2.31$ & $.025$ & Measure 5b(i); redundant alerts dropped \\
\quad registered coding (redundant dropped)\newline\textit{Alert; logistic GLMM} & Registered measure & $-1.25$ $[-1.81, -0.69]$ & -- & $1.3\times10^{-5}$ & Measure 5b(i) \\
First click correct, against the licensed-set benchmark\newline\textit{Person; paired $t$} & Additional & $+63.7$ pp $[60.8, 66.6]$ & $t(55.0) = 44.05$ & $1.4\times10^{-44}$ & -- \\
Time to correct click from alert onset (ms)\newline\textit{Person; paired $t$} & Additional & $-520$ & $t(55.0) = -5.33$ & $1.9\times10^{-6}$ & -- \\
Time to correct click from alert offset (ms)\newline\textit{Person; paired $t$} & Additional & $+620$ & $t(55.0) = 6.41$ & $3.5\times10^{-8}$ & -- \\
Answered mid-utterance: ms from onset\newline\textit{Person; paired $t$} & Additional & $-1236$ & $t(49.0) = -16.24$ & $2.4\times10^{-21}$ & -- \\
Share answered mid-utterance, underspecified vs fully specified\newline\textit{Alert; $\chi^2$} & Additional & 26.3\% vs 29.5\% & $\chi^2(1) = 2.42$ & $.120$ & -- \\
Prior belief in the first choice (per rating step)\newline\textit{Candidate cell; conditional logit} & Additional & OR $1.58$ $[1.36, 1.83]$ & $z = 6.01$ & $1.8\times10^{-9}$ & -- \\
\quad adding the prior to the display\newline\textit{Likelihood-ratio test} & Additional & -- & $\chi^2(1) = 36.81$ & $1.3\times10^{-9}$ & -- \\
Listener alignment: prior-using over literal\newline\textit{Trial (\PP{}); intercept-only LMM} & Registered exploratory & $+2.37$ $[0.65, 4.10]$ & -- & $.010$ & H3; margin (3) $-$ (2); no decision rule \\
Listener alignment: prior-using over no-update\newline\textit{Trial (\PP{}); intercept-only LMM} & Registered exploratory & $+0.19$ $[-0.44, 0.82]$ & -- & $.520$ & H3; margin (3) $-$ (1); no decision rule \\
Two-referent alerts: referents caught against one\newline\textit{Alert; one-sample $t$} & Additional & $+0.65$ & $t(164.0) = 14.30$ & $1.2\times10^{-30}$ & -- \\
\addlinespace[3pt]\multicolumn{6}{@{}l}{\textit{Allocation}} \\
Unbriefed events alerted of\newline\textit{Event; raw rates} & Registered exploratory & $+55.8$ pp $[53.3, 58.2]$ & -- & -- & What each agent leaves unsaid \\
Events with neither briefing nor alert\newline\textit{Event; raw rates} & Registered exploratory & $-27.1$ pp $[-28.6, -25.5]$ & -- & -- & What each agent leaves unsaid \\
Redundant fully specified alerts\newline\textit{Person; Welch $t$} & Registered exploratory & $-18.2$ pp $[-21.7, -14.7]$ & $t(72.9) = -10.31$ & $7.0\times10^{-16}$ & Redundancy rate \\
Late events caught (onset frame 4--5)\newline\textit{Person; Welch $t$} & Additional & $+39.6$ pp $[29.9, 49.3]$ & $t(100.5) = 8.09$ & $1.4\times10^{-12}$ & -- \\
\quad under low awareness\newline\textit{Person; Welch $t$} & Additional & $+62.6$ pp $[53.0, 72.1]$ & $t(81.2) = 13.03$ & $1.3\times10^{-21}$ & -- \\
The same 70 event types, caught\newline\textit{Event type; paired $t$} & Additional & $+7.4$ pp & $t(69.0) = 2.08$ & $.041$ & -- \\
First click after a beep on an unfound critical\newline\textit{Person; Welch $t$} & Additional & 71.7\% vs 43.3\% & $t(109.8) = 5.82$ & $6.0\times10^{-8}$ & -- \\
\addlinespace[3pt]\multicolumn{6}{@{}l}{\textit{Participant experience}} \\
Scenes rated helpful, \PP{} $-$ \Baseline{}\newline\textit{Trial; logistic GLMM (M4)} & Registered test & $+1.39$ $[0.32, 2.46]$ & $z = 2.54$ & $.011$ & H4b; two-sided; registered two-sided \\
Helpfulness, agent $\times$ trial position\newline\textit{M4 term} & Registered model term & $+0.082$ $[0.025, 0.140]$ & $z = 2.80$ & $.0050$ & H4b; drift across the session \\
Helpfulness, per-person share\newline\textit{Person; Welch $t$} & Additional & $+9.5$ pp $[2.6, 16.3]$ & $t(87.5) = 2.73$ & $.0076$ & -- \\
Helpfulness, per-person slopes (share per trial)\newline\textit{Person; Welch $t$} & Additional & $+0.0068$ $[0.0009, 0.0127]$ & $t(106.7) = 2.29$ & $.024$ & -- \\
Mental demand, \PP{} $-$ \Baseline{}\newline\textit{Person; linear model} & Registered test & $-0.39$ $[-1.03, 0.25]$ & -- & $.234$ & H4a; two-sided; registered two-sided \\
Nine further items (smallest $p$: told me what I needed)\newline\textit{Person; Welch $t$, BH family} & Registered test & $+0.57$ $[0.11, 1.03]$ & $t(111.2) = 2.44$ & $.016$ & Nine items; BH $q = .05$; $p_{\mathrm{BH}} = .148$; none survive \\
Ten-item composite (scale points)\newline\textit{Person; Welch $t$} & Additional (post hoc) & $+0.195$ $[-0.135, 0.524]$ & $t(113.2) = 1.17$ & $.244$ & Not summed in the registration \\
\end{longtable}
\vspace{-8pt}{\scriptsize $p$ is two-sided, as in the paper; where the registration specified a one-sided test, the last column gives its $p$ in the registered direction. Estimates are on the scale of the model named (log-odds, Brier points, ms or percentage points); follow-ups are simple effects of a registered model.\par}
\endgroup

\section{Models}

Every behavioural model has the form
\[ g\big(\mathbb{E}[y_i]\big) = \beta_0 + \beta_1\,\mathrm{agent}_i + \beta_2\,\mathrm{awareness}_i +
\beta_3\,(\mathrm{agent}\times\mathrm{awareness})_i + \beta_4\,\mathrm{complexity}_i +
\beta_5\,\mathrm{position}_i + u_{p(i)} + w_{s(i)}, \]
with crossed random intercepts for participants ($u_p$) and scenes ($w_s$). Detection, speed and helpfulness
use a logit link, speed as a discrete-time hazard on 2.25\,s intervals in which an event never caught is
censored at the end of the trial. Belief error is Gaussian and adds by-participant awareness slopes and
by-scene agent slopes; this full structure converged cleanly, so the registered simplification was not
needed. Factors are sum-coded ($\pm 1$; \PP{}, HA and four critical events at $+1$), so each coefficient is half
the difference between two levels; the alert-level decoding hazard (M5) instead compares its conditions with
\PP{}'s fully specified alerts. Models were fitted in R \rVersion{} with lme4 and lmerTest \cite{bates2015,kuznetsova2017}; Table~\ref{tab:fixed_short} gives their fixed effects.

\begingroup
\scriptsize
\begin{longtable}{L{5.2cm}rrlllL{2.6cm}}
\caption{Fixed effects of the fitted models.}\label{tab:fixed_short}\\
\toprule
Term & Estimate & SE & 95\% CI & Statistic & $p$ & exp(estimate) [95\% CI] \\
\midrule
\endfirsthead
\multicolumn{7}{@{}l}{\textit{Table~\ref{tab:fixed_short}, continued}}\\
\toprule
Term & Estimate & SE & 95\% CI & Statistic & $p$ & exp(estimate) [95\% CI] \\
\midrule
\endhead
\bottomrule
\endlastfoot
\addlinespace[3pt]\multicolumn{7}{@{}l}{\textit{M1: Critical event caught (event)}} \\
Intercept & $+1.830$ & $0.105$ & $[1.624, 2.036]$ & $z = 17.43$ & $5.3\times10^{-68}$ & $6.23$ $[5.07, 7.66]$ \\
Agent (\PP{} $=+1$) & $+0.274$ & $0.093$ & $[0.092, 0.456]$ & $z = 2.95$ & $.0032$ & $1.32$ $[1.10, 1.58]$ \\
Awareness (HA $=+1$) & $+0.116$ & $0.057$ & $[0.005, 0.228]$ & $z = 2.05$ & $.041$ & $1.12$ $[1.00, 1.26]$ \\
Complexity (four events $=+1$) & $-0.230$ & $0.057$ & $[-0.341, -0.119]$ & $z = -4.05$ & $5.1\times10^{-5}$ & $0.79$ $[0.71, 0.89]$ \\
Trial position (centred) & $+0.023$ & $0.005$ & $[0.012, 0.033]$ & $z = 4.10$ & $4.1\times10^{-5}$ & $1.02$ $[1.01, 1.03]$ \\
Agent $\times$ awareness & $-0.138$ & $0.032$ & $[-0.201, -0.076]$ & $z = -4.35$ & $1.4\times10^{-5}$ & $0.87$ $[0.82, 0.93]$ \\
\addlinespace[3pt]\multicolumn{7}{@{}l}{\textit{M2: Event caught in a 2.25\,s interval (event $\times$ interval at risk)}} \\
Intercept & $-1.540$ & $0.070$ & $[-1.677, -1.402]$ & $z = -21.90$ & $2.8\times10^{-106}$ & $0.21$ $[0.19, 0.25]$ \\
Agent (\PP{} $=+1$) & $+0.112$ & $0.048$ & $[0.018, 0.205]$ & $z = 2.35$ & $.019$ & $1.12$ $[1.02, 1.23]$ \\
Awareness (HA $=+1$) & $+0.108$ & $0.047$ & $[0.015, 0.200]$ & $z = 2.28$ & $.022$ & $1.11$ $[1.02, 1.22]$ \\
Complexity (four events $=+1$) & $-0.169$ & $0.047$ & $[-0.261, -0.076]$ & $z = -3.58$ & $3.5\times10^{-4}$ & $0.84$ $[0.77, 0.93]$ \\
Trial position (centred) & $+0.021$ & $0.003$ & $[0.016, 0.026]$ & $z = 8.01$ & $1.1\times10^{-15}$ & $1.02$ $[1.02, 1.03]$ \\
Agent $\times$ awareness & $-0.062$ & $0.015$ & $[-0.091, -0.033]$ & $z = -4.20$ & $2.6\times10^{-5}$ & $0.94$ $[0.91, 0.97]$ \\
\textit{Intervals 2--8: \Baseline{} hazard, 7 terms, not shown} &  &  &  &  &  &  \\
\addlinespace[3pt]\multicolumn{7}{@{}l}{\textit{M3: Post-trial belief error, Brier (trial)}} \\
Intercept & $+0.2851$ & $0.0208$ & $[0.2436, 0.3267]$ & $t(73.5) = 13.69$ & $6.7\times10^{-22}$ & -- \\
Agent (\PP{} $=+1$) & $-0.0102$ & $0.0172$ & $[-0.0443, 0.0238]$ & $t(115.3) = -0.60$ & $.552$ & -- \\
Awareness (HA $=+1$) & $-0.0225$ & $0.0136$ & $[-0.0511, 0.0061]$ & $t(17.9) = -1.65$ & $.116$ & -- \\
Complexity (four events $=+1$) & $+0.0315$ & $0.0110$ & $[0.0082, 0.0547]$ & $t(17.0) = 2.86$ & $.011$ & -- \\
Trial position (centred) & $-0.0089$ & $0.0008$ & $[-0.0104, -0.0074]$ & $t(2084.4) = -11.69$ & $1.3\times10^{-30}$ & -- \\
Agent $\times$ awareness & $+0.0144$ & $0.0068$ & $[0.0003, 0.0285]$ & $t(20.5) = 2.12$ & $.046$ & -- \\
\addlinespace[3pt]\multicolumn{7}{@{}l}{\textit{M4: Scene rated helpful (trial)}} \\
Intercept & $+3.734$ & $0.347$ & $[3.053, 4.415]$ & $z = 10.75$ & $6.1\times10^{-27}$ & $41.86$ $[21.19, 82.71]$ \\
Agent (\PP{} $=+1$) & $+0.694$ & $0.273$ & $[0.159, 1.230]$ & $z = 2.54$ & $.011$ & $2.00$ $[1.17, 3.42]$ \\
Trial position (centred) & $-0.029$ & $0.015$ & $[-0.057, 0.000]$ & $z = -1.94$ & $.052$ & $0.97$ $[0.94, 1.00]$ \\
Agent $\times$ trial position & $+0.041$ & $0.015$ & $[0.012, 0.070]$ & $z = 2.80$ & $.0050$ & $1.04$ $[1.01, 1.07]$ \\
\addlinespace[3pt]\multicolumn{7}{@{}l}{\textit{M5: Alert answered in a 1\,s bin (alert $\times$ bin at risk)}} \\
Intercept & $-3.001$ & $0.157$ & $[-3.308, -2.694]$ & $z = -19.17$ & $6.3\times10^{-82}$ & $0.05$ $[0.04, 0.07]$ \\
\PP{} underspecified (ref.\ \PP{} fully specified) & $+0.106$ & $0.065$ & $[-0.021, 0.232]$ & $z = 1.64$ & $.101$ & $1.11$ $[0.98, 1.26]$ \\
\BaselineCap{} fully specified (ref.\ \PP{} fully specified) & $+0.422$ & $0.127$ & $[0.174, 0.670]$ & $z = 3.33$ & $8.6\times10^{-4}$ & $1.53$ $[1.19, 1.95]$ \\
Awareness HA (ref.\ LA) & $+0.099$ & $0.132$ & $[-0.161, 0.358]$ & $z = 0.75$ & $.456$ & $1.10$ $[0.85, 1.43]$ \\
Four events (ref.\ three) & $-0.146$ & $0.132$ & $[-0.405, 0.113]$ & $z = -1.10$ & $.269$ & $0.86$ $[0.67, 1.12]$ \\
Trial position (centred) & $+0.023$ & $0.004$ & $[0.016, 0.030]$ & $z = 6.36$ & $2.0\times10^{-10}$ & $1.02$ $[1.02, 1.03]$ \\
\textit{Intervals 2--21: \Baseline{} hazard, 20 terms, not shown} &  &  &  &  &  &  \\
\addlinespace[3pt]\multicolumn{7}{@{}l}{\textit{M6: As M3, with agent $\times$ trial position}} \\
Intercept & $+0.2851$ & $0.0208$ & $[0.2435, 0.3266]$ & $t(72.3) = 13.68$ & $9.5\times10^{-22}$ & -- \\
Agent (\PP{} $=+1$) & $-0.0102$ & $0.0165$ & $[-0.0428, 0.0224]$ & $t(112.5) = -0.62$ & $.537$ & -- \\
Awareness (HA $=+1$) & $-0.0226$ & $0.0135$ & $[-0.0511, 0.0058]$ & $t(17.0) = -1.68$ & $.112$ & -- \\
Complexity (four events $=+1$) & $+0.0379$ & $0.0135$ & $[0.0095, 0.0664]$ & $t(17.0) = 2.81$ & $.012$ & -- \\
Trial position (centred) & $-0.0089$ & $0.0008$ & $[-0.0104, -0.0074]$ & $t(2093.9) = -11.64$ & $2.3\times10^{-30}$ & -- \\
Agent $\times$ awareness & $+0.0153$ & $0.0044$ & $[0.0067, 0.0239]$ & $t(2085.0) = 3.48$ & $5.0\times10^{-4}$ & -- \\
Agent $\times$ trial position & $-0.0020$ & $0.0008$ & $[-0.0035, -0.0006]$ & $t(2089.5) = -2.68$ & $.0073$ & -- \\
\addlinespace[3pt]\multicolumn{7}{@{}l}{\textit{M7: As M1, with agent $\times$ trial position}} \\
Intercept & $+1.830$ & $0.105$ & $[1.624, 2.036]$ & $z = 17.42$ & $5.4\times10^{-68}$ & $6.23$ $[5.07, 7.66]$ \\
Agent (\PP{} $=+1$) & $+0.273$ & $0.093$ & $[0.091, 0.456]$ & $z = 2.94$ & $.0033$ & $1.31$ $[1.10, 1.58]$ \\
Awareness (HA $=+1$) & $+0.116$ & $0.057$ & $[0.005, 0.228]$ & $z = 2.05$ & $.040$ & $1.12$ $[1.01, 1.26]$ \\
Complexity (four events $=+1$) & $-0.230$ & $0.057$ & $[-0.341, -0.119]$ & $z = -4.06$ & $5.0\times10^{-5}$ & $0.79$ $[0.71, 0.89]$ \\
Trial position (centred) & $+0.022$ & $0.006$ & $[0.011, 0.033]$ & $z = 3.99$ & $6.6\times10^{-5}$ & $1.02$ $[1.01, 1.03]$ \\
Agent $\times$ awareness & $-0.138$ & $0.032$ & $[-0.200, -0.075]$ & $z = -4.32$ & $1.6\times10^{-5}$ & $0.87$ $[0.82, 0.93]$ \\
Agent $\times$ trial position & $-0.002$ & $0.006$ & $[-0.013, 0.008]$ & $z = -0.43$ & $.670$ & $1.00$ $[0.99, 1.01]$ \\
\end{longtable}
\vspace{-8pt}{\scriptsize Log-odds for M1, M2, M4, M5 and M7; Brier points for M3 and M6. Gaussian models give $t$ with Satterthwaite degrees of freedom, logistic models Wald $z$. The exponentiated estimate is on the odds scale; \PP{} against \Baseline{} is the exponent of twice the agent estimate.\par}
\endgroup

\end{document}